# Orbital Hall conductivity and orbital diffusion length of Vanadium thin films by Hanle magnetoresistance

Montserrat X. Aguilar-Pujol[1,2], Isabel C. Arango[1], Eoin Dolan[1,2], You Ba[1], Marco Gobbi[3,4], Luis E. Hueso[1,4] and Fèlix Casanova[1,4,5,*]

[1]CIC NanoGUNE BRTA, 20018 Donostia-San Sebastian, Basque Country, Spain
[2]UPV-EHU, 20018 Donostia-San Sebastian, Basque Country, Spain
[3]Centro de Física de Materiales CSIC-UPV/EHU, 20018 Donostia-San Sebastian, Basque Country, Spain
[4]IKERBASQUE, Basque Foundation for Science, 48009 Bilbao, Basque Country, Spain
[5]Lead contact
*Correspondence: f.casanova@nanogune.eu

## SUMMARY

In spintronics, the spin Hall effect has been widely used to generate and detect spin currents in materials with strong spin-orbit coupling such as Pt and Ta. Recently, its orbital counterpart has drawn attention as a new tool to generate and detect orbital currents and thus investigate orbital transport parameters. In this study, we investigate vanadium (V), a 3d transition metal with weak spin-orbit coupling but with a theoretically large orbital Hall conductivity. We measure a large Hanle magnetoresistance in V thin films with a magnitude comparable to that of heavy metals and at least one order of magnitude higher than the spin Hall magnetoresistance observed in a $Y_3Fe_5O_{12}$/V bilayer, pointing to the orbital Hall origin of the effect. A fit of the magnetic-field dependence and thickness dependence of the Hanle magnetoresistance to the standard diffusion model allows us to quantify the orbital diffusion length (~2 nm) and the orbital Hall conductivity (~78 ($\hbar$/2e) $\Omega^{-1}$cm$^{-1}$) of V. The obtained orbital Hall conductivity is two orders of magnitude smaller than theoretical calculations of the intrinsic value, suggesting there is an important role of disorder.

## KEYWORDS

Orbital Hall effect, spin Hall effect, orbitronics, Hanle magnetoresistance, vanadium.

## INTRODUCTION

The spin Hall effect (SHE) converts a charge current into a spin current and, reciprocally, its inverse (ISHE) converts a spin current into a charge current. These two effects are thus widely used in spintronics to generate and detect spin currents without the use of a ferromagnetic material as the spin source[1,2]. Indeed, they are of utmost importance in novel applications such as the SHE to write magnetic elements in magnetic random-access memories (MRAMs) via spin-orbit torques[3–5] or the ISHE to readout magnetic elements in magnetoelectric spin-orbit (MESO) logic[5–7]. Since SHE depends on the strength of the spin-orbit coupling (SOC), which scales with the atomic number (Z) as $Z^4$ [8,9], SHE studies have mainly focused on heavy metals (HMs), essentially 4d and 5d transition metals[8,10–12]. Theoretical studies also proposed the existence of the orbital Hall effect (OHE), the orbital analogue of the SHE, in which electrons with opposite orbital angular momentum deflect in opposite directions when a charge current is applied through a conductor, giving rise to a transverse orbital current[13–17]. The OHE was firstly calculated in the well-known HMs[13,14] and, more recently, it has been extended to light metals (LMs) such as the 3d transition metals[15–17], which show weak SOC. According to these theoretical calculations, 3d transition metals present higher intrinsic orbital ($\sigma_{OH}^{int}$~ $10^3$–$10^4$ ($\hbar$/e) $\Omega^{-1}$cm$^{-1}$) than spin ($\sigma_{SH}^{int}$~ $10^2$ ($\hbar$/e) $\Omega^{-1}$cm$^{-1}$) Hall conductivities[16,17]; because of that, LMs have been considered ideal candidates to explore orbital physics. Experimental studies have quantified the orbital transport parameters of 3d transition metals by magneto-optical Kerr effect in Ti[18], Cr[19], and V[20], by spin-torque ferromagnetic resonance in Ti[21] and V[22], by electron magnetic circular dichroism also in Ti[23] or by Hanle Magnetoresistance (HMR) in Mn[24]. THz emission has also been used to study orbital transport in both 3d[25] and 5d transition metals[26].

HMR arises because of the combination of the SHE and ISHE together with the presence of an external magnetic field H producing a change in the resistivity and giving rise to the magnetoresistance effect[27]. When a charge current density $J_q$ is applied along the *x*-axis of a metal thin film, a spin angular momentum current along z-axis is created due to the SHE, with spin polarization along *y*. The spin currents are reflected back into the film at the top and bottom surfaces, inducing a spin accumulation near the surfaces and generating extra charge current ($J_q'$) due to the ISHE. This reduces the Drude resistivity of the thin film by a correction proportional to $\theta_{SH}^2$, where $\theta_{SH}$ is the spin Hall angle, when the thickness of the film is of the order of the spin diffusion length (see Figure 1A). By means of $H$, we can modify $J_q'$. If H is applied perpendicular to the spin polarization (*x*- or *z*-axis), we induce spin precession due to Hanle effect, changing the spin accumulation and diffusion (spin current reflection) and increasing the longitudinal resistivity $\rho_{L0}$ with $H$, eventually saturating at the Drude resistivity (see Figure 1B). If H is applied parallel to the initial spin polarization (*y*-axis), no precession will occur, keeping $\rho_{L0}$ to its reduced value. Because of the SHE requirement, it was originally explored on HMs[28–31]. Since the orbital angular momentum also precesses with $H$, HMR should also occur in the presence of OHE. Due to the important contribution of the OHE in LMs, it has been recently observed in Mn[24]. The phenomenon has also been theoretically investigated in two-dimensional materials, further extending the understanding of the orbital HMR[32].

In contrast, the spin Hall magnetoresistance (SMR)[33–35] arises when a HM is put in contact with a ferromagnet (FM). The spin current with spin polarization $s$, originated because of the SHE, interacts with the magnetization $M$ of the FM at the HM/FM interface, which can be reflected (when $M \parallel s$) or absorbed via spin transfer torque (when $M \perp s$). Although there is an orbital analogue of the SMR, the orbital Hall magnetoresistance[36] which arises due to the interaction of orbital currents with $M$ of the FM at a LM/FM interface via the orbital torque[37], it requires that the FM have a considerable orbital magnetization component[38].

In this work, we report a HMR of the order of $10^{-4}$ on sputtered V thin films. A patterned Hall bar allows to perform longitudinal and transverse magnetotransport measurements by rotating the sample along different axis and sweeping the magnetic field up to 9 T. We observe the symmetry expected for HMR in the temperature range from 25 K to 300 K in samples of different thicknesses ranging from 4 nm to 30 nm and different resistivities. From the analysis of the SMR and spin-mixing conductance of $Y_3Fe_5O_{12}$/V samples, we confirm the orbital Hall origin of the observed HMR and quantify the orbital transport parameters of V.

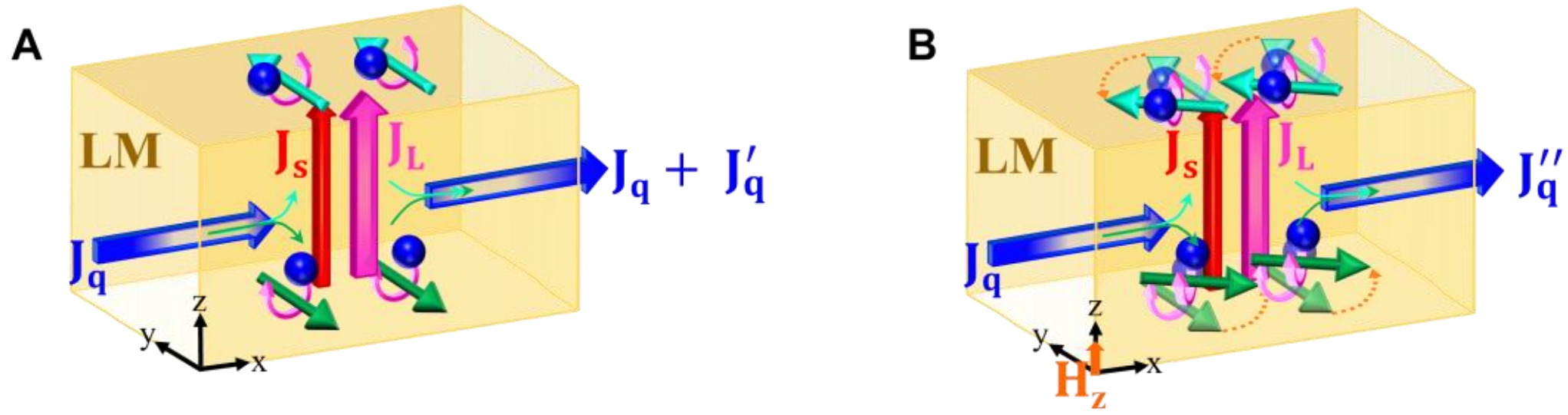

***Figure 1.*** **Schematic illustration of the Hanle magnetoresistance.** (A) Influence of the SHE and OHE and their inverse effects in the decrease of the resistivity of a LM thin film, when there is no precession in absence of magnetic field $H$ or with $H$ along the spin/orbital polarization (y-direction). (B) The spin and orbital angular momenta accumulated at the edges precess due to Hanle effect around a perpendicular $H$ (z- or x-direction, in the sketch, the former example is shown), reducing the extra current generated by the ISHE and inverse OHE and, thus, increasing the resistivity of the LM thin film.

## RESULTS

According to the spin (and orbit) diffusion theory, the resistivity of the NM layer in the longitudinal ($\rho_L$) and transverse ($\rho_T$) configuration changes depending on the orientation of an applied $H$ for the HMR or on the orientation of the magnetization of an adjacent FM for the SMR. This results in a characteristic dependence of $\rho_L$ and $\rho_T$ of the film that vary according to the equations:

$$\rho_L^{HMR,SMR} = \rho_{L0} + \Delta\rho_1^{HMR,SMR}\left(1 - n_y^2\right) \quad \text{(Equation 1)}$$

$$\rho_T^{HMR,SMR} = \Delta\rho_1^{HMR,SMR} n_x n_y + \Delta\rho_2^{HMR,SMR} n_z \quad \text{(Equation 2)}$$

where $\boldsymbol{n}$ denotes the orientation of the magnetization $M$ in the FM film for the SMR ($\boldsymbol{n} = \boldsymbol{M}/M$), or the orientation of the external magnetic field for the HMR ($\boldsymbol{n} = \boldsymbol{H}/H$). $\Delta\rho_1^{HMR,SMR}$ is the amplitude of the HMR or SMR and $\Delta\rho_2^{HMR,SMR}$ accounts for an anomalous Hall-like contribution due to HMR or SMR[28,33,34,39,40].

ADMR and FDMR measurements are taken in the system Si/$SiO_2$/V($t_V$ nm)/$SiO_2$(5 nm) or Si/$SiO_2$/V($t_V$ nm)/SiN(8 nm) (see methods for more details), where $t_V$ denotes the thickness, for samples with a resistivity of $\rho_{L0}$~ 270 μΩ·cm (see Figure S1). Although both capping and substrate materials act as diffusion barriers, the presence of an interfacial vanadium oxide layer cannot be fully excluded. Such a layer could slightly reduce the effective conducting thickness, but it is not expected to influence the overall results.

Figure 2 shows the measurement configuration (Figure 2A) and representative results of the longitudinal ADMR (Figure 2B), longitudinal FDMR (Figure 2C) and transverse FDMR (Figure 2D) measurements taken in a 6-nm-thick film at $T$ = 100 K. Both ADMR and FDMR show the shape and symmetry expected by HMR between 25 K and 300 K (see the temperature dependence of the ADMR in Figure S2). The result of the longitudinal ADMR is plotted as $\Delta\rho_L/\rho_{L0} = \Delta R_L/R_L = [R_L(\alpha,\beta) - R_L(90°)]/R_L(90°)$, where $R_L = V_L/I$ is the longitudinal resistance (see Figure 2A). We observe the typical $\cos^2(\alpha,\beta)$ modulation related to the HMR geometry[28], which gives an amplitude of the order of $10^{-4}$ in both $\alpha$ and $\beta$ planes (Figure 2B). Similar values have been found in Pt[28–30]. A small difference of the order of $10^{-5}$ is present between the amplitude of $\alpha$ and $\beta$ plane, corresponding to a $\sin^2\gamma$ modulation in $\gamma$ plane which cannot be due to HMR, for which the modulation should vanish (see also Note S2). The origin of such extra magnetoresistance, which is negative when $H$ is applied along *z* (see the difference between longitudinal FDMR along *x* and along *z* in Figure 2C), is most likely related to the weak localization which appears in disordered systems and is characterized by a resistivity increase below a characteristic crossover temperature[41,42], which is present in all our samples (see Figure S1A).

The longitudinal FDMR (Figure 2C), plotted as $\Delta\rho_L/\rho_{L0} = \Delta R_L/R_L = [R_L(H_i) - R_L(H=0)]/R_L(H=0)$, with *i=x,y,z* shows that the amplitude depends on the strength of the magnetic field with a parabolic increase at low fields ($\propto H_{x,z}^2$) that tends to flatten at high fields, with similar values along the *x*- and *z*-axes (except for the small variation due to weak localization mentioned above). In contrast, a small positive FDMR is found along the *y*-axis, which we ascribe to the in-plane weak localization contribution[43]. Figure 2D, plotted as $\rho_T/\rho_{L0} = R_T(H_z)l/(wR_L)$, where $R_T = V_T/I$ is the transverse resistance, shows the normalized transverse resistivity where the contribution from the ordinary Hall effect is subtracted by performing a linear fit of the transverse FDMR as shown in Figure S3 and detailed in Note S3. We see a strong dependence on the magnetic field only when it is applied out of plane, with a linear increase with magnetic field ($\propto H_z$) at low fields and tending to saturation above ~3 T. Overall, the observed signatures of the measured longitudinal ADMR, longitudinal FDMR and transverse FDMR confirm the presence of HMR in our V thin films.

Furthermore, we perform ADMR and FDMR measurements in a set of samples with same thickness but different resistivities. In the range of resistivities evaluated ($\rho_{L0}$~ 171–464 μΩ·cm), all of them showed the characteristic HMR signatures discussed above. Among them, we find that the amplitude of the HMR increases with resistivity, with values one order of magnitude higher for highly resistive samples with the same thickness (see Figure S4 and Note S4 for more details).

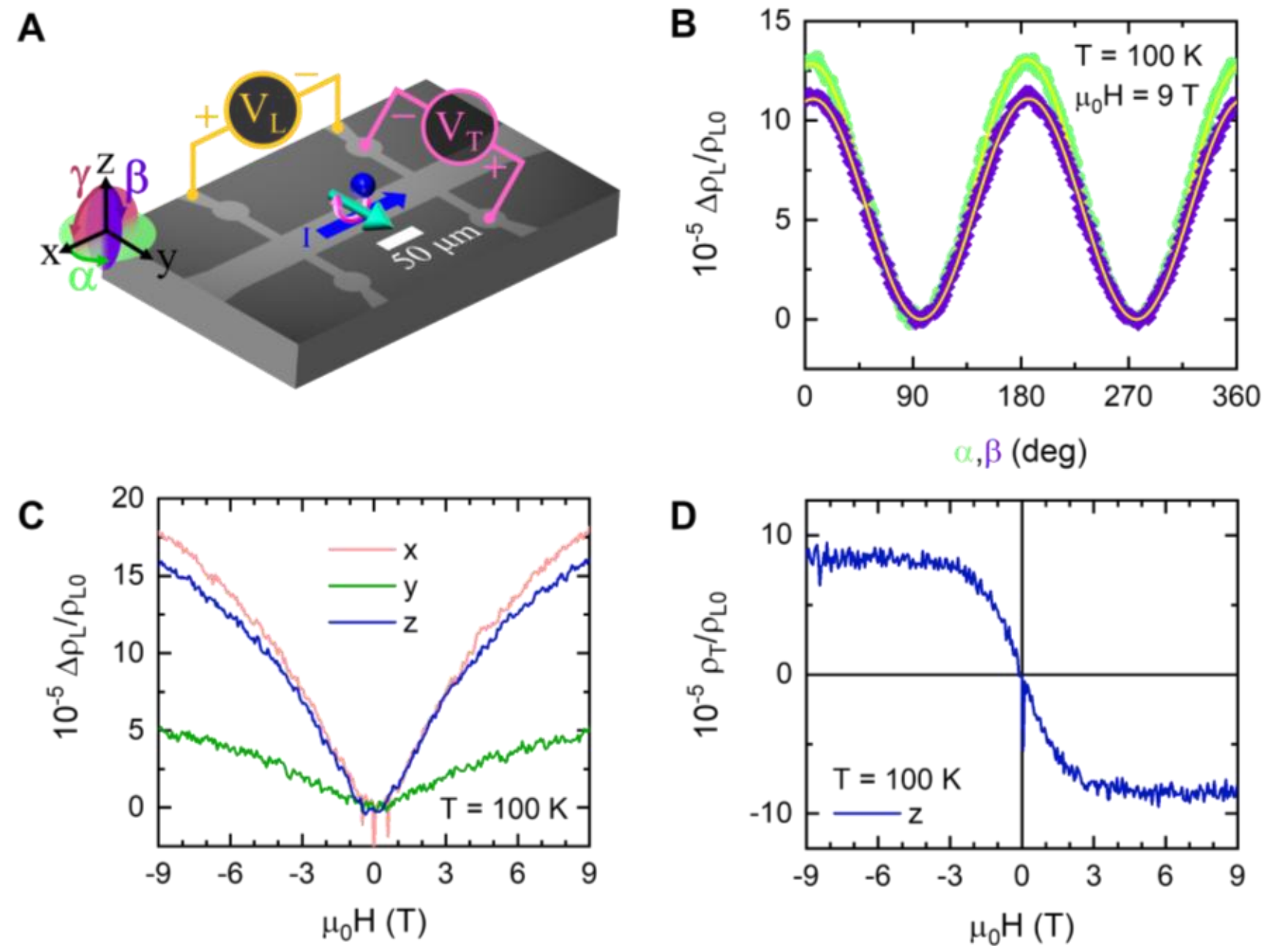

***Figure 2.*** **HMR measurements.** (A) Optical microscope image of a vanadium Hall bar showing the experimental setup for HMR and SMR measurements. A charge current $I$ is applied along x-direction in the Hall bar and then the longitudinal $V_L$ and transverse $V_T$ voltages are measured as depicted in the image. The three $H$-rotation planes $(\alpha, \beta, \gamma)$ are illustrated. (B) Longitudinal ADMR measurement at $T$ =100 K and $\mu_0 H$ = 9 T in $\alpha, \beta$ planes. Yellow lines correspond to a $cos^2(\alpha, \beta)$ fit. (C) Longitudinal and (D) transverse FDMR measurements performed at 100 K with $H$ applied along the three main axes. Data corresponds to sample Si/$SiO_2$/V(6 nm)/$SiO_2$(5 nm). See also Figures S2, S3 and S4.

Although the HMR can originate from both the SHE and the OHE, we expect that, given the weak SOC in V, it arises mostly from the OHE and not from the SHE. Indeed, theoretical calculations for bcc V predict a large $\sigma_{OH}$ from 4500 to 6050 ($\hbar$/e) $\Omega^{-1}cm^{-1}$ with a much smaller $\sigma_{SH}$ from –13 to –90 ($\hbar$/e) $\Omega^{-1}cm^{-1}$ [16,17,44,45]. Importantly, HMR provides a way to quantify angular momentum transport parameters, which in HMs correspond to spin[28,29], whereas in LMs are mostly orbital[24]. In principle, orbital interfacial effects could also contribute to the HMR, but disentangling them from the bulk would be very challenging. For that, we come back to Equations (1) and (2) and focus on $\Delta\rho_1^{HMR}$ and $\Delta\rho_2^{HMR}$, which we can express as a function of the orbital Hall angle $\theta_{OH}$ and the orbital diffusion length $\lambda_{OD}$:

$$\Delta\rho_1^{HMR} = 2\rho_{L0}\theta_{OH}^2\left\{\frac{\lambda_{OD}}{t_{LM}}tanh\left(\frac{t_{LM}}{2\lambda_{OD}}\right) - Re\left[\frac{\Lambda}{t_{LM}}tanh\left(\frac{t_{LM}}{2\Lambda}\right)\right]\right\} \qquad \text{(Equation 3)}$$

$$\Delta\rho_2^{HMR} = 2\rho_{L0}\theta_{OH}^2 Im\left[\frac{\Lambda}{t_{LM}}tanh\left(\frac{t_{LM}}{2\Lambda}\right)\right] \qquad \text{(Equation 4)}$$

where $t_{LM}$ is the film thickness of the light metal, and $\Lambda^{-1} = \sqrt{\frac{1}{\lambda_{OD}^2} + i\frac{g\mu_B B}{D_O\hbar}}$ with $g$ the gyromagnetic factor, $\mu_B$ is the Bohr magneton, $D_O$ the orbital diffusion coefficient, and $\hbar$ is the reduced Planck constant[28,29].

To extract the 3 unknown parameters ($\theta_{OH}$, $\lambda_{OD}$, and $D_O$), we can simultaneously fit the longitudinal FDMR to Equation 3 and transverse FDMR to Equation 4, but the use of 3 free fitting parameters with 2 data sets gives correlations between the free parameters. In order to obtain an extra data set, the HMR was measured in samples with different thickness in the range between 4 and 30 nm on Si/$SiO_2$ substrates (see results in Figures 3A and 3C for 100 K and 290 K, respectively). For that, we chose a set of samples where $\rho_{L0}$ does not change significantly with thickness (see Figure S1). By using Equation 3, we were able to fit the thickness dependence of the HMR amplitude, obtained from the longitudinal ADMR measurement in $\alpha$-plane (such as the one shown in Figure 2B), at a fixed magnetic field. In this fit, $\lambda_{OD}$ is uncoupled from the other two parameters, since it is essentially constrained by the position of the observed maximum but not by the amplitude. By fixing the obtained value of $\lambda_{OD}^{V}$, we perform again the simultaneous fit of the longitudinal and transverse FDMR in a 6-nm-thick film (see Figures 3B and 3D). For that, we subtract the contribution from the weak localization in the longitudinal FDMR by doing $\Delta\rho_{L'}/\rho_{L0} = \Delta\rho_{L}^{x}/\rho_{L0} - \Delta\rho_{L,T}^{y}/\rho_{L0}$. This allows us to reliably extract the other two free parameters, $\theta_{OH}$ and $D_{O}$. Additionally, the orbital relaxation time $\tau_{OD} = \lambda_{OD}^{2}/D_{O}$ and the orbital Hall conductivity $\sigma_{OH} = \theta_{OH}/\rho_{L0}$ can also be derived. The orbital transport parameters of V extracted from the fittings are listed in Table I for two specific temperatures, with similar values at 100 K and 290 K ($\lambda_{OD}^{V}$~2 nm and $\theta_{OH}^{V}$~0.020).

Moreover, we also analyze the impact of resistivity on the orbital transport parameters of the V thin films. To do so, we perform a simultaneous triple-variable fit of the thickness-dependent ADMR signal of high-resistivity samples and the longitudinal and transverse FDMR datasets of a 6-nm-thick V sample with a resistivity of $\rho_{L0}$~ 378 μΩ·cm, at 290 K (for more details, see Note S5 and Figure S5). While most of the transport parameters remain unchanged within the error (see Table I), $\theta_{OH}$ and $\sigma_{OH}$ are affected by the change of resistivity. $\theta_{OH}$ increases by ~14% with respect to the previous set of samples with $\rho_{L0}$~ 270 μΩ·cm, while $\sigma_{OH}$ decreases by ~19%, an interesting result that is discussed below.

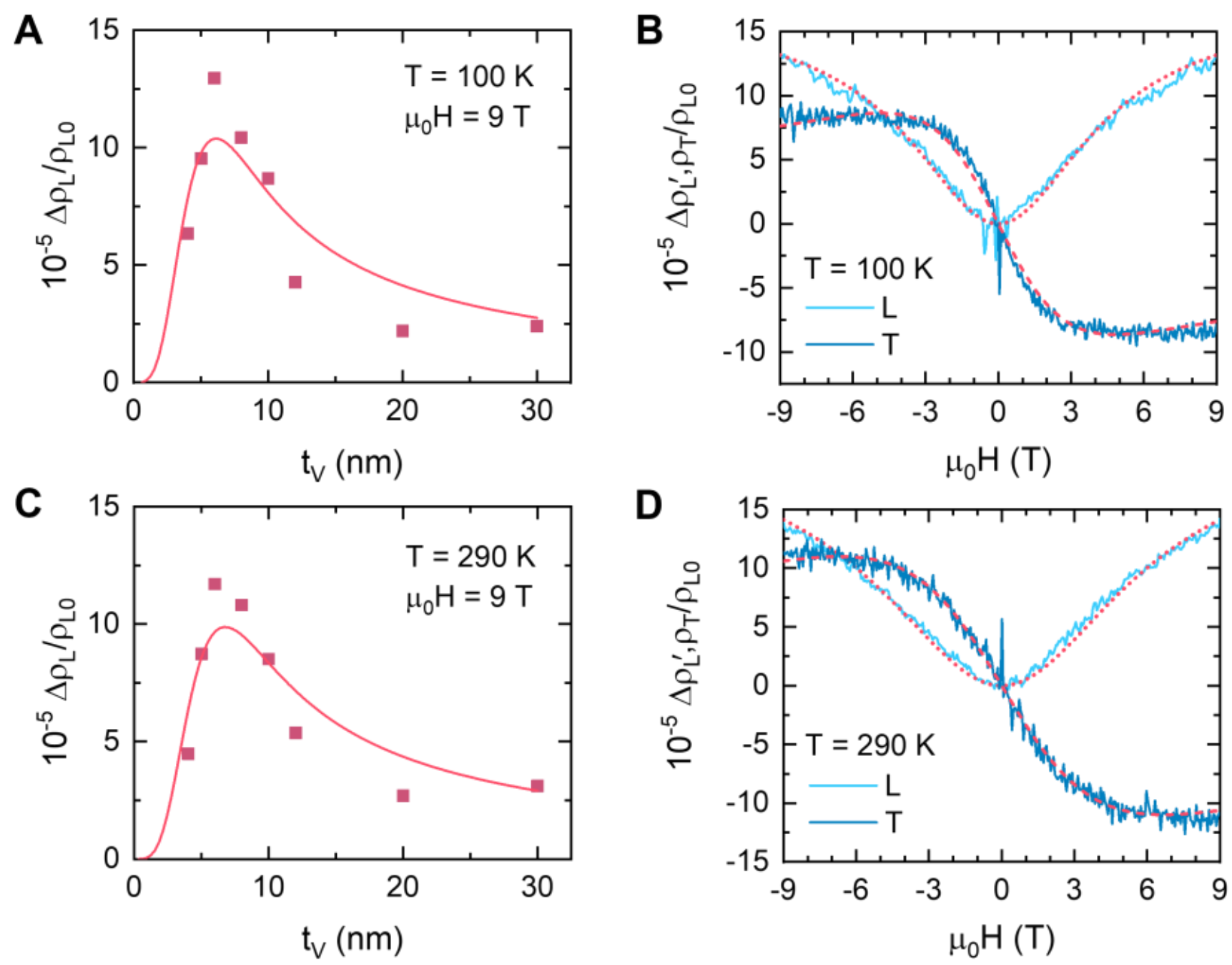

**Figure 3. HMR thickness and magnetic field dependence in V thin films with fittings to the HMR equations.** (A) and (C) HMR amplitude at 9 T extracted from the longitudinal ADMR measurements in $\alpha$ plane (pink solid squares) as a function of thickness at 100 K and 290 K, respectively, with their corresponding fittings of Equation 3 (red solid line). (B) and (D) Longitudinal ($L$) (light blue line) and transverse ($T$) (dark blue line) FDMR for sample Si/$SiO_2$/V(6 nm)/$SiO_2$(5 nm) at 100 K and 290 K, respectively, with their corresponding fittings of Equation 3 (dotted lines) and Equation 4 (dashed lines). See also Figures S5 and S8.

**Table 1. Orbital transport parameters of V extracted from the fittings at 100 K and 290 K.**

| $T$ (K) | $\rho_{L0}$ (μΩ·cm) | $\lambda_{OD}^{V}$ (nm) | $\theta_{OH}^{V}$ | $D_{O}^{V}$ (mm²/s) | $\tau_{O}^{V}$ (ps) | $\sigma_{OH}^{V}$ [(ħ/2e) $\Omega^{-1}cm^{-1}$] |
|---|---|---|---|---|---|---|
| 100 | 264.4 | 1.8 ± 0.3 | 0.020 ± 0.001 | 1.3 ± 0.4 | 2.5 ± 0.2 | 76 ± 4 |
| 290 | 270.3 | 2.1 ± 0.3 | 0.021 ± 0.001 | 2.2 ± 0.6 | 2.0 ± 0.2 | 78 ± 4 |

| 290 | 378.1 | 2.1 ± 0.3 | 0.024 ± 0.001 | 2.0 ± 0.6 | 2.2 ± 0.2 | 63 ± 3 |
|---|---|---|---|---|---|---|

In order to know whether there is any contribution from the SHE in the HMR, we can perform SMR measurements using YIG as an adjacent ferromagnetic insulator. Since the SMR is modulated by the magnetization of YIG, which can only interact with spin (but not orbital) currents[38], its amplitude will depend on the SHE only. The measurements performed in YIG/V(7 nm)/$SiO_2$(5 nm) at 100 K are represented in Figure 4, where we plot in-plane transverse measurements (equivalent to the longitudinal ones, compare second term in Equation 1 with first term in Equation 2) because they give a larger signal-to-noise ratio in the otherwise small signal. Figure 4A shows the normalized transverse FDMR plotted as $\Delta\rho_T/\rho_{L0} = \Delta R_T l/(wR_L) = [R_T(H) - R_T(90°)]l/[wR_L(90°)]$, with the magnetic field applied in plane at $\alpha$ = 45º and $\alpha$ = 135º, where the occurrence of the magnetization reversal of YIG is clearly detected around 0.3 mT. The gap between the two curves accounts for the magnitude of the SMR, which is of the order of $10^{-6}$, two orders of magnitude lower that in the HMR case. The same amplitude appears in the normalized transverse ADMR measurements, $\Delta\rho_T/\rho_{L0} = \Delta R_T l/(wR_L) = [R_T(\alpha) - R_T(90°)]l/[wR_L(90°)]$, at low magnetic fields (20 mT) where a sin $\alpha$·cos $\alpha$ modulation is present as shown in Figure 4B and expected from the SMR theory (see Equation 2). At high magnetic fields, where the HMR may appear, we see again the sin $\alpha$·cos $\alpha$ modulation but with an amplitude more than one order of magnitude higher, although reduced as compared with Si/$SiO_2$/V samples of similar thickness (see Figures 3A and S6, and Note S6 for more details). The presence of a clear SMR signal confirms that we have spin currents generated by SHE, but it is more than one order of magnitude lower than in YIG/Pt[28,46], in which the SMR amplitude and the HMR amplitude at 9 T have been reported to be of the same order.

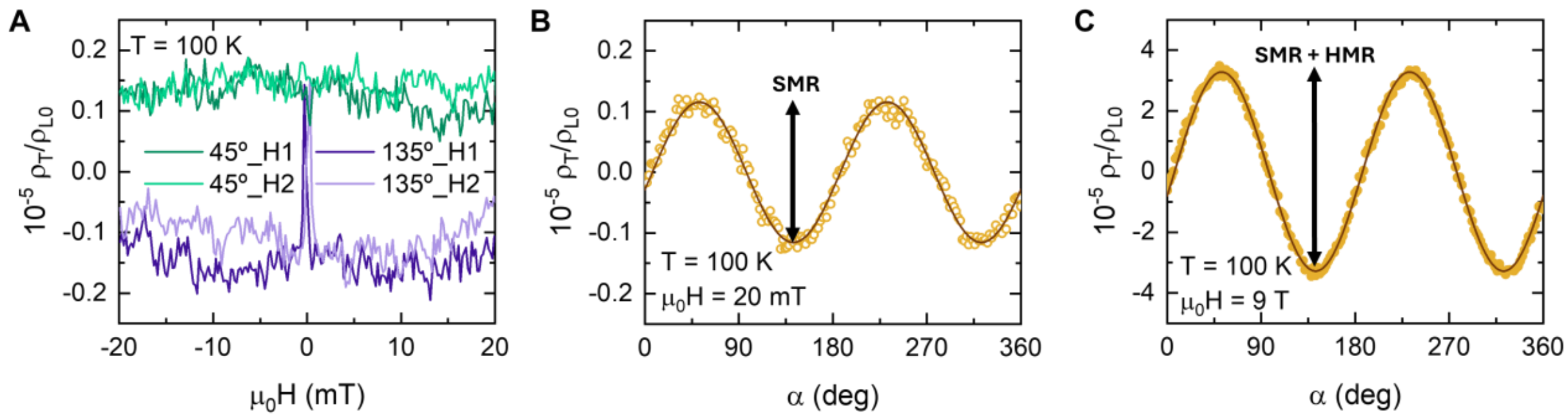

**Figure 4. SMR measurements on YIG/V/$SiO_2$.** (A) Normalized transverse FDMR measurements performed at 45º and 135º in a-plane in a low magnetic field range. (B) Normalized transverse ADMR measurements at $\mu_0 H$ = 20 mT, where the brown line corresponds to a sin $\alpha$·cos $\alpha$ fit. The arrow indicates the SMR amplitude $\Delta\rho_1^{SMR}/\rho_{L0}$. (C) Normalized transverse ADMR measurements at $\mu_0 H$ = 9 T. Brown line corresponds to a sin $\alpha$·cos $\alpha$ fit. The arrow indicates the total amplitude, which contains the HMR ($\Delta\rho_1^{HMR}/\rho_{L0}$) and SMR contributions. Data correspond to measurements performed at 100 K on sample YIG/V(7 nm)/$SiO_2$(5 nm). See also Figure S6.

Whereas the much smaller amplitude of SMR as compared to HMR in YIG/V seem to confirm that HMR is dominated by the OHE contribution, there is still the possibility that a strongly reduced spin-mixing conductance ($g^{\uparrow\downarrow}$) at the YIG/V interface explains the different HMR and SMR amplitudes solely with the presence of SHE. In order to rule out this possibility, we performed FMR on YIG/V samples. FMR allows to extract the real part of the interfacial spin-mixing conductance, $g_r^{\uparrow\downarrow}$, which determines the ability of the interface to transmit spins. From the Gilbert damping constants of the YIG single layer, $\alpha_{YIG}$, and YIG/V bilayer, $\alpha_{YIG/V}$, we can obtain $g_r^{\uparrow\downarrow}$:

$$g_r^{\uparrow\downarrow} = \frac{4\pi M_S t_{YIG}}{g\mu_B}\left(\alpha_{YIG/NM} - \alpha_{YIG}\right) \quad \text{(Equation 5)}$$

where $M_S$, $t_{YIG}$, $g$ and $\mu_B$ are the saturation magnetization, YIG thickness (318 nm), Landé factor and Bohr magneton, respectively. By performing a Lorentz fit of each FMR spectrum (see dotted lines in Figure 5A), we determined $M_S$ and $\alpha$ from the frequency dependencies of the resonant field $H_{res}$ and linewidth $\Delta H$, respectively.

The frequency dependence of $H_{res}$ follows the Kittel formula[47], and at $T$ = 100 K we found an effective magnetization $4\pi M_{eff}$ = 2673 ± 1 G, which we can consider that equals the saturation magnetization[48,49]. $\alpha$ can be obtained from the linear increase of $\Delta H$ with frequency, $f$, following:

$$\Delta H = \Delta H_0 + \frac{4\pi\alpha f}{\gamma} \qquad \text{(Equation 6)}$$

being $\Delta H_0$ the inhomogeneous broadening and $\gamma$ the gyromagnetic ratio. Figure 5B shows the linear increase of $\Delta H$ with frequency, for a YIG film and a YIG/V(20nm)/SiO2(5nm) film, in the frequency range between 10 and 19 GHz. From the linear fit of the data, the damping obtained is $\alpha_{YIG}$ = (1.9 ± 0.1)×10$^{-4}$ and $\alpha_{YIG/V}$ = (1.0 ± 0.1)×10$^{-3}$ at $T$ = 100 K. By using Equation 5, we calculate $g_r^{\uparrow\downarrow}$ = (4.0 ± 0.3)×10$^{19}$ m$^{-2}$ at 100 K, one order of magnitude larger than previously reported at 300 K in YIG/V using 20-nm-thick YIG films[50].

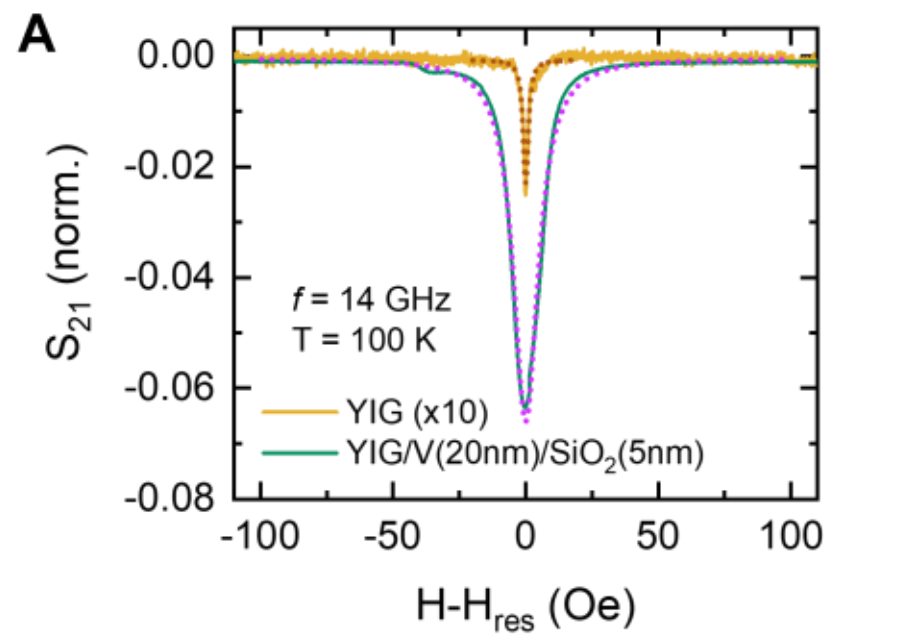

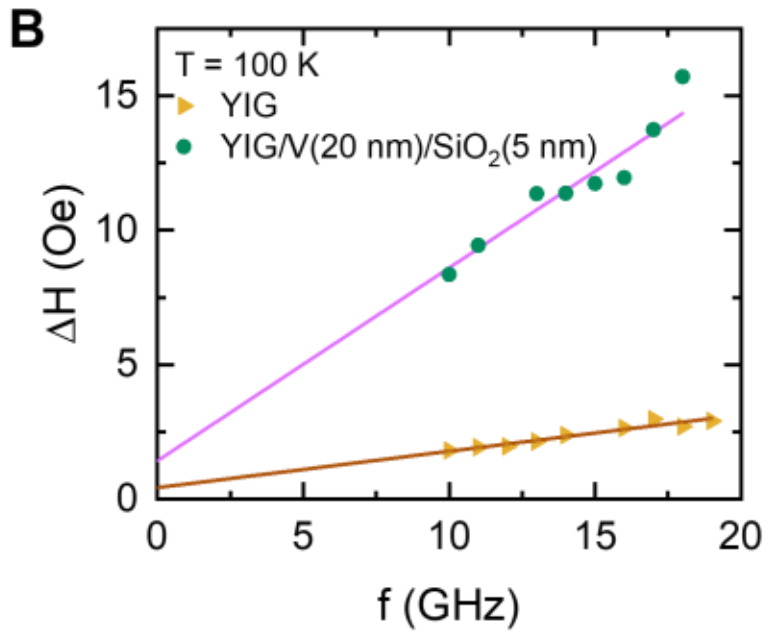

**Figure 5. FMR on YIG and YIG/V/SiO$_2$ samples.** (A) Experimental FMR spectra of a YIG film and a YIG/V(20 nm)/SiO$_2$(5 nm) film, measured at $T$ = 100 K and $f$ = 14 GHz. Dotted lines represent the fittings to a Lorentzian curve. (B) Frequency dependencies of FMR linewidth of a YIG film and a YIG/V(20 nm)/SiO$_2$(5 nm) film, at $T$ = 100 K. Solid lines correspond to a linear fit of the experimental data to Equation 6.

With the extracted value of $g_r^{\uparrow\downarrow}$, and considering the transport parameters shown in Table I to be of spin origin, we calculate the expected amplitude of the SMR under this hypothesis. The amplitude $\Delta\rho_1^{SMR}$ can be expressed in terms of the spin transport parameters as[51]:

$$\frac{\Delta\rho_1^{SMR}}{\rho_{L0}} \approx \theta_{SH}^2 \frac{\lambda_{SD}}{t_{LM}} \frac{2\lambda_{SD} G_r^{\uparrow\downarrow} \tanh^2 \frac{t_{LM}}{2\lambda_{SD}}}{\frac{1}{\rho_{L0}} + 2\lambda_{SD} G_r^{\uparrow\downarrow} \coth \frac{t_{LM}}{\lambda_{SD}}} \qquad \text{(Equation 7)}$$

where $\theta_{SH}$ is the spin Hall angle, $\lambda_{SD}$ is the spin diffusion length, and $G_r^{\uparrow\downarrow} = e^2 g_r^{\uparrow\downarrow}/h$, with $e$ the elementary charge and $h$ the Panck constant. It should be noted that Equation 7 assumes minimal interface spin-orbit coupling contributions, focusing on bulk spin-orbit effects[51]. By replacing the values of Table I (low resistivity samples at $T$ = 100 K) in Equation 7, we get $\Delta\rho_1^{SMR}/\rho_{L0}$ = 7.9×10$^{-5}$, which is two orders of magnitude larger than what we obtain experimentally (see Figure 4B). Since the transport parameters obtained from HMR cannot explain the observed amplitude of the SMR, we can confirm that they are not associated with spin angular momentum but with the orbital angular momentum. In contrast, we can use the SMR amplitude to estimate the $\sigma_{SH}^V$ value. Taking the spin diffusion length of V ($\lambda_{SD}^V$), which has been determined experimentally to be $\lambda_{SD}^V$ ~ 16 nm for samples with similar resistivity[50,52], and $G_r^{\uparrow\downarrow}$ at the YIG/V interface, obtained from spin pumping, we can use Equation 7 to estimate the absolute value of $\sigma_{SH}$. Our estimation, ~24 (ℏ/e) Ω$^{-1}$cm$^{-1}$, lies within the range of theoretical –13 to –90 (ℏ/e) Ω$^{-1}$cm$^{-1}$ values[16,17,44,45], further evidencing the spin origin of the SMR.

## DISCUSSION

Taking this into account, we will discuss the orbital transport parameters obtained from HMR. We find that they are in the same range as the ones reported for Mn with the same technique[24], with $\theta_{OH}^V$ being also similar to the value reported for Ti, $\theta_{OH}^{Ti}$ = 0.015 ± 0.002, by MOKE[18]. If we look at $\sigma_{OH}^V$, its magnitude is 2 orders lower than the theoretical value of the intrinsic orbital Hall conductivity, $\sigma_{OH}^{int}$ ~ 4500 – 6050 (ℏ/e) Ω$^{-1}$cm$^{-1}$[16,17,44,45], a large difference that was also found for Mn[24] and might be explained by the large disorder in our sputtered thin films (see Note S7). The first theoretical works showed that, while the value of $\sigma_{OH}^{int}$ is constant in the moderately dirty regime and does not depend on $\rho$, $\sigma_{OH}^{int}$ is predicted to decrease approximately proportional to $\rho^{-2}$ in the high-resistivity regime with large disorder[13], similar to the behaviour for $\sigma_{SH}$, in which these two regimes have been observed experimentally[53,54]. Although we observe a decrease of conductivity from $\sigma_{OH}^{V}$ = 78 ± 4 (ℏ/2e) $\Omega^{-1}$cm$^{-1}$ to $\sigma_{OH}^{V}$ = 63 ± 3 (ℏ/e) $\Omega^{-1}$cm$^{-1}$ when resistivity increases from $\rho_{L0}$~ 270 µΩ·cm to 380 µΩ·cm, it does not follow the $\rho^{-2}$ trend. This suggests that we are at the stage where $\sigma_{OH}^{int}$ starts to decrease but this does not justify the low $\sigma_{OH}^{V}$ as compared with theoretical predictions. The large discrepancy existing between theoretical predictions and our measured $\sigma_{OH}$ may stem from neglecting disorder-induced vertex corrections, which a recent work[55] shows can strongly suppress the $\sigma_{OH}$ in certain cases. Furthermore, a newly introduced quantum-kinetic theory of electronic angular-momentum linear response[56,57] restricts the $\sigma_{OH}$ to intraband contributions. In nonmagnetic, centrosymmetric light metals such as vanadium, this implies an intrinsic $\sigma_{OH}$ that scales at least quadratically with the weak SOC, in contrast to conventional SOC-independent formulas, and hence naturally yields much smaller $\sigma_{OH}$ values. A report on OHE in vanadium with lower resistivity ($\rho_{L0}$ ~ 41 µΩ·cm) by MOKE, quantifies both $\sigma_{OH}$ and $\lambda_{OD}$, but the strong correlation between the two values prevents an independent quantification[20].

We next consider the orbital diffusion constant $D_O$, which provides complementary insight into the orbital transport dynamics, and we compare it to the charge diffusion constant $D_q$, which can be estimated using the Einstein relation $D_q = 1/\left(e^2\rho N(E_F)\right)$. Here, $e$ is the elementary charge and $N(E_F)$ is the density of states at the Fermi level. Using the reported value for V, $N(E_F)$ ~ 28×10$^{22}$ eV$^{-1}$cm$^{-3}$[58], we obtain $D_q^V$ ~ 8 mm$^2$/s for the samples with lower resistivity ($\rho_{L0}$ ~ 270 µΩ·cm). This value is four to six times larger than $D_O^V$ extracted from the HMR fittings at the same sample (see Table I). A difference between $D_q$ and $D_O$ has also been observed in Ti[18] and is in fact expected, according to theoretical studies[18,59]. In particular, the general theory of diffusion developed by X. Ning *et al.*[59] predicts that, while charge and spin diffusion constants are typically of similar magnitude, the orbital diffusion constant can be significantly lower, highlighting the distinct dynamics of orbital transport.

Interestingly, we find that $D_O$ exhibits a more complex dependence on temperature and resistivity than initially expected. Specifically, a comparison of the same sample measured at 100 K and 290 K shows a clear increase in $D_O$ with temperature, as given in Table I, despite only a small change in resistivity because of the disordered metal behaviour. On the contrary, when comparing samples with significantly different resistivities (low- and high-resistivity datasets), at 290 K, we obtain nearly identical $D_O$ values. To further investigate the relationship between $D_O$, resistivity and temperature, and confirm the robustness of the previous result, we performed an additional fitting on an 8-nm-thick film from the low-resistivity set (see Note S8 for more details). The obtained orbital transport parameters are consistent with those from the 6-nm-thick sample (see Table S2). These results collectively suggest that $D_O$ is more sensitive to temperature than to resistivity, which is predominantly determined by disorder. Although the origin of this behaviour remains unclear, it may motivate future theoretical studies.

Regarding the orbital diffusion length, we extract $\lambda_{OD}^{V}$ ~ 2 nm which does not depend on the resistivity, suggesting it is robust against disorder. The obtained value is similar to the reported one by spin-torque ferromagnetic resonance $\lambda_{OD}^{V}$ = 3.6 nm[22]. $\lambda_{OD}^{V}$ is also similar to the one reported for Cr, $\lambda_{OD}^{Cr}$ = 6.6 ± 0.6 nm, by MOKE[19] and to the one for Mn, $\lambda_{OD}^{Mn}$~ 2 nm, by HMR[24]. According to recent first-principles scattering calculations by M. Rang *et al.*, the orbital current decays rapidly, resulting in $\lambda_{OD}$ of less than 1 nm, with $\lambda_{OD}^{V}$ = 0.5 nm[60] for the specific case of V, and with no influence of disorder. They suggest that the longer length scales being reported experimentally may arise because orbital current is converted into a spin current in a few atomic layers and what is being observed is the resulting spin diffusion length ($\lambda_{SD}$) after orbital to spin current conversion. Experimental $\lambda_{SD}$ values for V samples with similar resistivity are reported to be $\lambda_{SD}^{V}$ ~ 16 nm [50,52]. Thus, our value $\lambda_{OD}^{V}$ ~ 2 nm lies in between the theoretical orbital diffusion length and the experimental spin diffusion length, consistent with the previous hypothesis. Further studies are required to clarify the origin of the measured orbital transport parameters and the huge spread in the orbital diffusion lengths already reported[18–25,36].

In summary, we have studied the HMR and SMR effects in V thin films. We find a large HMR amplitude of the order of 10 in the range of thickness explored from 4 nm to 30 nm, while the SMR amplitude is one to two orders of magnitude smaller. These results, in combination with FMR in YIG/V, can be explained by the fact that V is a light metal with an expected large orbital Hall but small spin Hall contribution, i.e., the measured HMR arises mostly from the OHE. Furthermore, by fitting our results to the HMR equations, we are able to quantify the orbital transport parameters. We determine an orbital Hall conductivity $\sigma_{OH}^{V} \sim 78$ ($\hbar$/2e) $\Omega^{-1}$cm$^{-1}$ that slightly decreases when resistivity increases from $\rho_{L0} \sim 270$ to ~ 380 μΩ·cm, while the orbital diffusion length $\lambda_{OD}^{V} \sim 2$ nm, remains independent of the resistivity in the same range. The large discrepancy between our experimentally obtained orbital Hall conductivity and the intrinsic one theoretically calculated may be related to the grade of disorder in our V thin films, which is not always fully considered in theoretical studies. Moreover, our study reveals a clear difference between charge and orbital diffusivities, a discrepancy that highlights the fundamentally different transport dynamics of orbital angular momentum. Our analysis also evidences that the orbital diffusion length of V is shorter than its spin diffusion length, but larger than its mean free path. Further investigation is needed to clarify the impact of disorder on the suppression of orbital transport, particularly its role in limiting the magnitude of both the orbital Hall conductivity and the orbital diffusion length observed in disordered thin films.

## METHODS

V samples were prepared by patterning a Hall bar with a length of l = 200 μm and a width of w = 50 μm on top of Si/$SiO_2$(150 nm) or $Y_3Fe_5O_{12}$ (yttrium iron garnet, YIG) grown on gadolinium gallium garnet (GGG). The samples were prepared by photolithography process and magnetron-sputtering deposition of V (50 W dc; 3 mTorr of Ar), followed by a capping layer of either 5 nm of $SiO_2$ (200 W rf; 3 mTorr of Ar) or 8 nm of SiN (100 W rf; 3 mTorr of Ar) to prevent oxidation.

Magnetotransport measurements were taken at a Physical Property Measurement System (PPMS) covering the temperature range 25 K $\leq T \leq$ 300 K. We performed angular-dependent magnetoresistance (ADMR) measurements and field-dependent magnetoresistance (FDMR) measurements by applying magnetic fields $H$ up to 9 T. For the ADMR, $H$ is rotated along 3 different planes: $xy$-plane, where $H$ is rotated by an angle $\alpha$ from the direction of the applied current ($x$), $yz$-plane, where $H$ is rotated by an angle $\beta$ from the out-of-plane direction ($z$), and $xz$-plane, where $H$ is rotated by an angle $\gamma$ from the out-of-plane direction ($z$) (see Figure 2A). For the FDMR, H is applied along the three main axes ($x$, $y$ and $z$). As sketched in Figure 2A, a 10 μA dc electric current is applied along the Hall bar with a Keithley 6221 current source meter, and the voltage is measured with a Keithley 2182 nanovoltmeter in the longitudinal $V_L$ or transverse $V_T$ configuration using a dc-reversal mode.

The magnetic properties of the samples have been studied with the ferromagnetic resonance (FMR) option of the PPMS, using a vector network analyser (VNA) connected with a coplanar wave guide, covering the temperature range 100 K $\leq T \leq$ 290 K. FMR measurements were performed at constant temperature and fixed microwave frequency (in the range of 10-19 GHz) while sweeping the magnetic field.

## RESOURCE AVAILABILITY

### *Lead contact*

Requests for further information and resources should be directed to and will be fulfilled by the lead contact, Fèlix Casanova (f.casanova@nanogune.eu).

### *Materials availability*

This study did not generate new unique materials.

### *Data and code availability*

- All data reported in this paper will be shared by the lead contact upon request.
- This paper does not report original code.

- Any additional information required to reanalyze the data reported in this paper is available from the lead contact upon request.

## ACKNOWLEDGMENTS

The authors thank Hyun-Woo Lee, Thierry Valet and Aurélien Manchon for fruitful discussions and acknowledge technical and human support provided by SGIker Medidas Magneticas Gipuzkoa (UPV/EHU/ERDF,EU). This work is supported by the Spanish MICIU/AEI/10.13039/501100011033 (Grant No. CEX2020-001038-M), by MICIU/AEI and ERDF/EU (Project No. PID2021-122511OB-I00), and by the European Union's Horizon 2020 research and innovation programme under Marie Skłodowska-Curie Grant Agreement No. 766025. M.X.A.-P. thanks the Spanish MICIU/AEI for a Ph.D. fellowship (grant No. PRE-2019-089833).

## AUTHOR CONTRIBUTIONS

Conceptualization, F.C.; methodology, M.X.A.P., E.D., and F.C.; investigation, M.X.A.P., I.C.A., and Y.B.; software, E.D.; writing—original draft, M.X.A.P.; writing—review & editing, F.C.; funding acquisition, M.G., F.C., and L.E.H.; supervision, M.G., L.E.H., and F.C.

## DECLARATION OF INTERESTS

The authors declare no competing interests.

## REFERENCES

1. Sinova, J., Valenzuela, S.O., Wunderlich, J., Back, C.H., and Jungwirth, T. (2015). Spin Hall effects. Rev Mod Phys *87*, 1213–1260. https://doi.org/10.1103/RevModPhys.87.1213.

2. Hirsch, J.E. (1999). Spin Hall Effect. Phys Rev Lett *83*, 1834–1837. https://doi.org/10.1103/PhysRevLett.83.1834.

3. Manchon, A., Železný, J., Miron, I.M., Jungwirth, T., Sinova, J., Thiaville, A., Garello, K., and Gambardella, P. (2019). Current-induced spin-orbit torques in ferromagnetic and antiferromagnetic systems. Rev Mod Phys *91*, 035004. https://doi.org/10.1103/RevModPhys.91.035004.

4. Shao, Q., Li, P., Liu, L., Yang, H., Fukami, S., Razavi, A., Wu, H., Wang, K., Freimuth, F., Mokrousov, Y., et al. (2021). Roadmap of Spin-Orbit Torques. IEEE Trans Magn *57*, 1–39. https://doi.org/10.1109/TMAG.2021.3078583.

5. Fert, A., Ramesh, R., Garcia, V., Casanova, F., and Bibes, M. (2024). Electrical control of magnetism by electric field and current-induced torques. Rev Mod Phys *96*, 015005. https://doi.org/10.1103/RevModPhys.96.015005.

6. Vaz, D.C., Lin, C.C., Plombon, J.J., Choi, W.Y., Groen, I., Arango, I.C., Chuvilin, A., Hueso, L.E., Nikonov, D.E., Li, H., et al. (2024). Voltage-based magnetization switching and reading in magnetoelectric spin-orbit nanodevices. Nat Commun *15*, 1902. https://doi.org/10.1038/s41467-024-45868-x.

7. Incorvia, J.A.C., Xiao, T.P., Zogbi, N., Naeemi, A., Adelmann, C., Catthoor, F., Tahoori, M., Casanova, F., Becherer, M., Prenat, G., et al. (2024). Spintronics for achieving system-level energy-efficient logic. Nature Reviews Electrical Engineering *1*, 700–713. https://doi.org/10.1038/s44287-024-00103-z.

8. Wang, H.L., Du, C.H., Pu, Y., Adur, R., Hammel, P.C., and Yang, F.Y. (2014). Scaling of spin hall angle in 3d, 4d, and 5d metals from Y3Fe5 O12 /metal spin pumping. Phys Rev Lett *112*, 197201. https://doi.org/10.1103/PhysRevLett.112.197201.

9. Sarma, D.D. (1981). Nature of dependence of spin-orbit splittings on atomic number. Proceedings of the Indian Academy of Sciences - Chemical Sciences *90*, 19–26. https://doi.org/10.1007/BF02841324.

10. Niimi, Y., and Otani, Y. (2015). Reciprocal spin Hall effects in conductors with strong spin-orbit coupling: A review. Reports on Progress in Physics *78*, 124501. https://doi.org/10.1088/0034-4885/78/12/124501.

11. Mosendz, O., Vlaminck, V., Pearson, J.E., Fradin, F.Y., Bauer, G.E.W., Bader, S.D., and Hoffmann, A. (2010). Detection and quantification of inverse spin Hall effect from spin pumping in permalloy/normal metal bilayers. Phys Rev B Condens Matter Mater Phys *82*, 214403. https://doi.org/10.1103/PhysRevB.82.214403.

12. Morota, M., Niimi, Y., Ohnishi, K., Wei, D.H., Tanaka, T., Kontani, H., Kimura, T., and Otani, Y. (2011). Indication of intrinsic spin Hall effect in 4d and 5d transition metals. Phys Rev B Condens Matter Mater Phys *83*, 174405. https://doi.org/10.1103/PhysRevB.83.174405.

13. Tanaka, T., Kontani, H., Naito, M., Naito, T., Hirashima, D.S., Yamada, K., and Inoue, J. (2008). Intrinsic spin Hall effect and orbital Hall effect in 4d and 5d transition metals. Phys Rev B Condens Matter Mater Phys *77*, 165117. https://doi.org/10.1103/PhysRevB.77.165117.

14. Kontani, H., Tanaka, T., Hirashima, D.S., Yamada, K., and Inoue, J. (2009). Giant orbital hall effect in transition metals: Origin of large spin and anomalous hall effects. Phys Rev Lett *102*, 016601. https://doi.org/10.1103/PhysRevLett.102.016601.

15. Go, D., Jo, D., Kim, C., and Lee, H.W. (2018). Intrinsic Spin and Orbital Hall Effects from Orbital Texture. Phys Rev Lett *121*, 086602. https://doi.org/10.1103/PhysRevLett.121.086602.

16. Jo, D., Go, D., and Lee, H.W. (2018). Gigantic intrinsic orbital Hall effects in weakly spin-orbit coupled metals. Phys Rev B *98*, 214405. https://doi.org/10.1103/PhysRevB.98.214405.

17. Salemi, L., and Oppeneer, P.M. (2022). First-principles theory of intrinsic spin and orbital Hall and Nernst effects in metallic monoatomic crystals. Phys Rev Mater *6*, 095001. https://doi.org/10.1103/PhysRevMaterials.6.095001.

18. Choi, Y.G., Jo, D., Ko, K.H., Go, D., Kim, K.H., Park, H.G., Kim, C., Min, B.C., Choi, G.M., and Lee, H.W. (2023). Observation of the orbital Hall effect in a light metal Ti. Nature *619*, 52–56. https://doi.org/10.1038/s41586-023-06101-9.

19. Lyalin, I., Alikhah, S., Berritta, M., Oppeneer, P.M., and Kawakami, R.K. (2023). Magneto-Optical Detection of the Orbital Hall Effect in Chromium. Phys Rev Lett *131*, 156702. https://doi.org/10.1103/PhysRevLett.131.156702.

20. Marui, Y., Kawaguchi, M., Sumi, S., Awano, H., Nakamura, K., and Hayashi, M. (2023). Spin and orbital Hall currents detected via current-induced magneto-optical Kerr effect in V and Pt. Phys Rev B *108*, 144436. https://doi.org/10.1103/PhysRevB.108.144436.

21. Hayashi, H., Jo, D., Go, D., Gao, T., Haku, S., Mokrousov, Y., Lee, H.W., and Ando, K. (2023). Observation of long-range orbital transport and giant orbital torque. Commun Phys *6*, 32. https://doi.org/10.1038/s42005-023-01139-7.

22. Liu, X., Liu, F., and Jiang, C. (2024). Harnessing orbital Hall effect for efficient orbital torque in light metal vanadium. J Magn Magn Mater *610*, 172585. https://doi.org/10.1016/j.jmmm.2024.172585.

23. Idrobo, J.C., Rusz, J., Datt, G., Jo, D., Alikhah, S., Muradas, D., Noumbe, U., Kamalakar, M. V., and P. M. Oppeneer, P.M. (2024). Direct observation of nanometer-scale orbital angular momentum accumulation. arXiv:2403.09269.

24. Sala, G., Wang, H., Legrand, W., and Gambardella, P. (2023). Orbital Hanle Magnetoresistance in a 3d Transition Metal. Phys Rev Lett *131*, 156703. https://doi.org/10.1103/PhysRevLett.131.156703.

25. Xu, Y., Zhang, F., Fert, A., Jaffres, H.Y., Liu, Y., Xu, R., Jiang, Y., Cheng, H., and Zhao, W. (2024). Orbitronics: light-induced orbital currents in Ni studied by terahertz emission experiments. Nat Commun *15*, 2043. https://doi.org/10.1038/s41467-024-46405-6.

26. Seifert, T.S., Go, D., Hayashi, H., Rouzegar, R., Freimuth, F., Ando, K., Mokrousov, Y., and Kampfrath, T. (2023). Time-domain observation of ballistic orbital-angular-momentum currents with giant relaxation length in tungsten. Nat Nanotechnol *18*, 1132–1138. https://doi.org/10.1038/s41565-023-01470-8.

27. Dyakonov, M.I. (2007). Magnetoresistance due to edge spin accumulation. Phys Rev Lett *99*, 126601. https://doi.org/10.1103/PhysRevLett.99.126601.

28. Vélez, S., Golovach, V.N., Bedoya-Pinto, A., Isasa, M., Sagasta, E., Abadia, M., Rogero, C., Hueso, L.E., Bergeret, F.S., and Casanova, F. (2016). Hanle Magnetoresistance in Thin Metal Films with Strong Spin-Orbit Coupling. Phys Rev Lett *116*, 016603. https://doi.org/10.1103/PhysRevLett.116.016603.

29. Li, J., Comstock, A.H., Sun, D., and Xu, X. (2022). Comprehensive demonstration of spin Hall Hanle effects in epitaxial Pt thin films. Phys Rev B *106*, 184420. https://doi.org/10.1103/PhysRevB.106.184420.

30. Maruyama, Y., Ohshima, R., Shigematsu, E., Ando, Y., and Shiraishi, M. (2023). Modulation of Hanle magnetoresistance in an ultrathin platinum film by ionic gating. Applied Physics Express *16*, 023004. https://doi.org/10.35848/1882-0786/acbc0a.

31. Wu, H., Zhang, X., Wan, C.H., Tao, B.S., Huang, L., Kong, W.J., and Han, X.F. (2016). Hanle magnetoresistance: The role of edge spin accumulation and interfacial spin current. Phys Rev B *94*, 174407. https://doi.org/10.1103/PhysRevB.94.174407.

32. Sun, H., and Vignale, G. (2025). Orbital magnetic moment dynamics and Hanle magnetoresistance in multilayered two-dimensional materials. Phys Rev B *111*, L180408. https://doi.org/10.1103/PhysRevB.111.L180408.

33. Chen, Y.T., Takahashi, S., Nakayama, H., Althammer, M., Goennenwein, S.T.B., Saitoh, E., and Bauer, G.E.W. (2013). Theory of spin Hall magnetoresistance. Phys Rev B *87*, 144411. https://doi.org/10.1103/PhysRevB.87.144411.

34. Zhang, X.P., Bergeret, F.S., and Golovach, V.N. (2019). Theory of Spin Hall Magnetoresistance from a Microscopic Perspective. Nano Lett *19*, 6330–6337. https://doi.org/10.1021/acs.nanolett.9b02459.

35. Althammer, M., Meyer, S., Nakayama, H., Schreier, M., Altmannshofer, S., Weiler, M., Huebl, H., Geprägs, S., Opel, M., Gross, R., et al. (2013). Quantitative study of the spin Hall magnetoresistance in ferromagnetic insulator/normal metal hybrids. Phys Rev B *87*, 224401. https://doi.org/10.1103/PhysRevB.87.224401.

36. Hayashi, H., and Ando, K. (2023). Orbital Hall magnetoresistance in Ni/Ti bilayers. Appl Phys Lett *123*, 172401. https://doi.org/10.1063/5.0170654.

37. Lee, D., Go, D., Park, H.J., Jeong, W., Ko, H.W., Yun, D., Jo, D., Lee, S., Go, G., Oh, J.H., et al. (2021). Orbital torque in magnetic bilayers. Nat Commun *12*, 6710. https://doi.org/10.1038/s41467-021-26650-9.

38. Go, D., Jo, D., Lee, H.W., Kläui, M., and Mokrousov, Y. (2021). Orbitronics: Orbital currents in solids. EPL *135*, 37001. https://doi.org/10.1209/0295-5075/ac2653.

39. Avci, C.O., Quindeau, A., Pai, C.F., Mann, M., Caretta, L., Tang, A.S., Onbasli, M.C., Ross, C.A., and Beach, G.S.D. (2017). Current-induced switching in a magnetic insulator. Nat Mater *16*, 309–314. https://doi.org/10.1038/nmat4812.

40. Husain, S., Figueiredo-Prestes, N., Fayet, O., Collin, S., Godel, F., Jacquet, E., Reyren, N., Jaffrès, H., and George, J.M. (2023). Origin of the anomalous Hall effect at the magnetic insulator/heavy metals interface. Appl Phys Lett *122*, 062403. https://doi.org/10.1063/5.0132895.

41. Bergmann, G. (1984). Weak localization in thin films. a time-of-flight experiment with conduction electrons. Phys Rep *107*, 1–58. https://doi.org/10.1016/0370-1573(84)90103-0.

42. Barone, C., Romeo, F., Pagano, S., Attanasio, C., Carapella, G., Cirillo, C., Galdi, A., Grimaldi, G., Guarino, A., Leo, A., et al. (2015). Nonequilibrium fluctuations as a distinctive feature of weak localization. Sci Rep *5*, 10705. https://doi.org/10.1038/srep10705.

43. Maekawa, S., and Fukuyama, H. (1981). Magnetoresistance in Two-Dimensional Disordered Systems: Effects of Zeeman Splitting and Spin-Orbit Scattering. J Physical Soc Japan *50*, 2516–2524. https://doi.org/10.1143/JPSJ.50.2516.

44. Go, D., Lee, H.-W., Oppeneer, P.M., Blügel, S., and Mokrousov, Y. (2024). First-principles calculation of orbital Hall effect by Wannier interpolation: Role of orbital dependence of the anomalous position. Phys Rev B *109*, 174435. https://doi.org/10.1103/PhysRevB.109.174435.

45. Rang, M., and Kelly, P.J. (2025). Orbital Hall effect in transition metals from first-principles scattering calculations. Phys Rev B *111*, 125121. https://doi.org/10.1103/PhysRevB.111.125121.

46. Hahn, C., De Loubens, G., Klein, O., Viret, M., Naletov, V. V., and Ben Youssef, J. (2013). Comparative measurements of inverse spin Hall effects and magnetoresistance in YIG/Pt and YIG/Ta. Phys Rev B Condens Matter Mater Phys *87*, 174417. https://doi.org/10.1103/PhysRevB.87.174417.

47. Kittel, C. (1948). On the theory of ferromagnetic resonance absorption. Physical Review *73*, 155. https://doi.org/10.1103/PhysRev.73.155.

48. Yang, F., and Chris Hammel, P. (2018). FMR-driven spin pumping in Y3Fe5O12-based structures. J Phys D Appl Phys *51*, 253001. https://doi.org/10.1088/1361-6463/aac249.

49. Jungfleisch, M.B., Chumak, A. V., Kehlberger, A., Lauer, V., Kim, D.H., Onbasli, M.C., Ross, C.A., Kläui, M., and Hillebrands, B. (2015). Thickness and power dependence of the spin-pumping effect in Y3Fe5 O12 /Pt heterostructures measured by the inverse spin Hall effect. Phys Rev B Condens Matter Mater Phys *91*, 134407. https://doi.org/10.1103/PhysRevB.91.134407.

50. Du, C., Wang, H., Yang, F., and Hammel, P.C. (2014). Systematic variation of spin-orbit coupling with d -orbital filling: Large inverse spin Hall effect in 3d transition metals. Phys Rev B Condens Matter Mater Phys *90*, 140407. https://doi.org/10.1103/PhysRevB.90.140407.

51. Chen, Y.T., Takahashi, S., Nakayama, H., Althammer, M., Goennenwein, S.T.B., Saitoh, E., and Bauer, G.E.W. (2016). Theory of spin Hall magnetoresistance (SMR) and related phenomena. Journal of Physics Condensed Matter *28*, 103004. https://doi.org/10.1088/0953-8984/28/10/103004.

52. Wang, T., Wang, W., Xie, Y., Warsi, M.A., Wu, J., Chen, Y., Lorenz, V.O., Fan, X., and Xiao, J.Q. (2017). Large spin Hall angle in vanadium film. Sci Rep *7*, 1306. https://doi.org/10.1038/s41598-017-01112-9.

53. Dushenko, S., Hokazono, M., Nakamura, K., Ando, Y., Shinjo, T., and Shiraishi, M. (2018). Tunable inverse spin Hall effect in nanometer-thick platinum films by ionic gating. Nat Commun *9*, 3118. https://doi.org/10.1038/s41467-018-05611-9.

54. Sagasta, E., Omori, Y., Isasa, M., Gradhand, M., Hueso, L.E., Niimi, Y., Otani, Y., and Casanova, F. (2016). Tuning the spin Hall effect of Pt from the moderately dirty to the superclean regime. Phys Rev B *94*, 060412. https://doi.org/10.1103/PhysRevB.94.060412.

55. Tang, P., and Bauer, G.E.W. (2024). Role of Disorder in the Intrinsic Orbital Hall Effect. Phys Rev Lett *133*, 186302. https://doi.org/10.1103/PhysRevLett.133.186302.

56. Valet, T., and Raimondi, R. (2025). Quantum kinetic theory of the linear response for weakly disordered multiband systems. Phys Rev B *111*, L041118. https://doi.org/10.1103/PhysRevB.111.L041118.

57. Valet, T., Jaffrès, H., Cros, V., and Raimondi, R. (2025). Quantum Kinetic Anatomy of Electron Angular Momenta Edge Accumulation. arXiv:2507.06771.

58. Palenskis, V., and Žitkevičius, E. (2018). Phonon Mediated Electron-Electron Scattering in Metals. World Journal of Condensed Matter Physics *08*, 115. https://doi.org/10.4236/wjcmp.2018.83008.

59. Ning, X., Pezo, A., Kim, K.-W., Zhao, W., Lee, K.-J., and Manchon, A. (2025). Orbital Diffusion, Polarization, and Swapping in Centrosymmetric Metals. Phys Rev Lett *134*, 026303. https://doi.org/10.1103/PhysRevLett.134.026303.

60. Rang, M., and Kelly, P.J. (2024). Orbital relaxation length from first-principles scattering calculations. Phys Rev B *109*, 214427. https://doi.org/10.1103/PhysRevB.109.214427.

## SUPPLEMENTAL INFORMATION

### Note S1. Resistivity of V thin films

We fabricated vanadium (V) thin films by sputtering following the deposition conditions specified in the main text. By changing the pre-sputtering time, resistivities ranging between $\rho_{L0}\sim$ 171–464 μΩ·cm at 290 K were obtained. We chose the set of samples shown in Figure S1A in order to have similar resistivities with thicknesses up to 30 nm. The resistivity of the set of V samples remains in the range $\rho_{L0}\sim$256–278 μΩ·cm (at 100 K) for $t_V \geq 6$ nm (Figure S1A). The samples with thickness $t_V \geq 6$ nm show a metallic behavior with a resistance minimum below 100 K that changes with thickness, indicating a disordered thin film. A characteristic resistivity curve as a function of temperature is shown in Figure S1C, for a 6-nm-thick film and in Figure S1D, for a 30-nm-thick film. The resistivity obtained for samples with $t_V < 6$ nm was always higher, showing an increase with decreasing temperature in the whole temperature range (10-300 K), characteristic of a more disordered metal than thicker films (Figure S1B). Despite this, we saw an important decrease in the amplitude of the HMR as expected for a diffusive phenomenon (see Figures 3A and 3C in the main text).

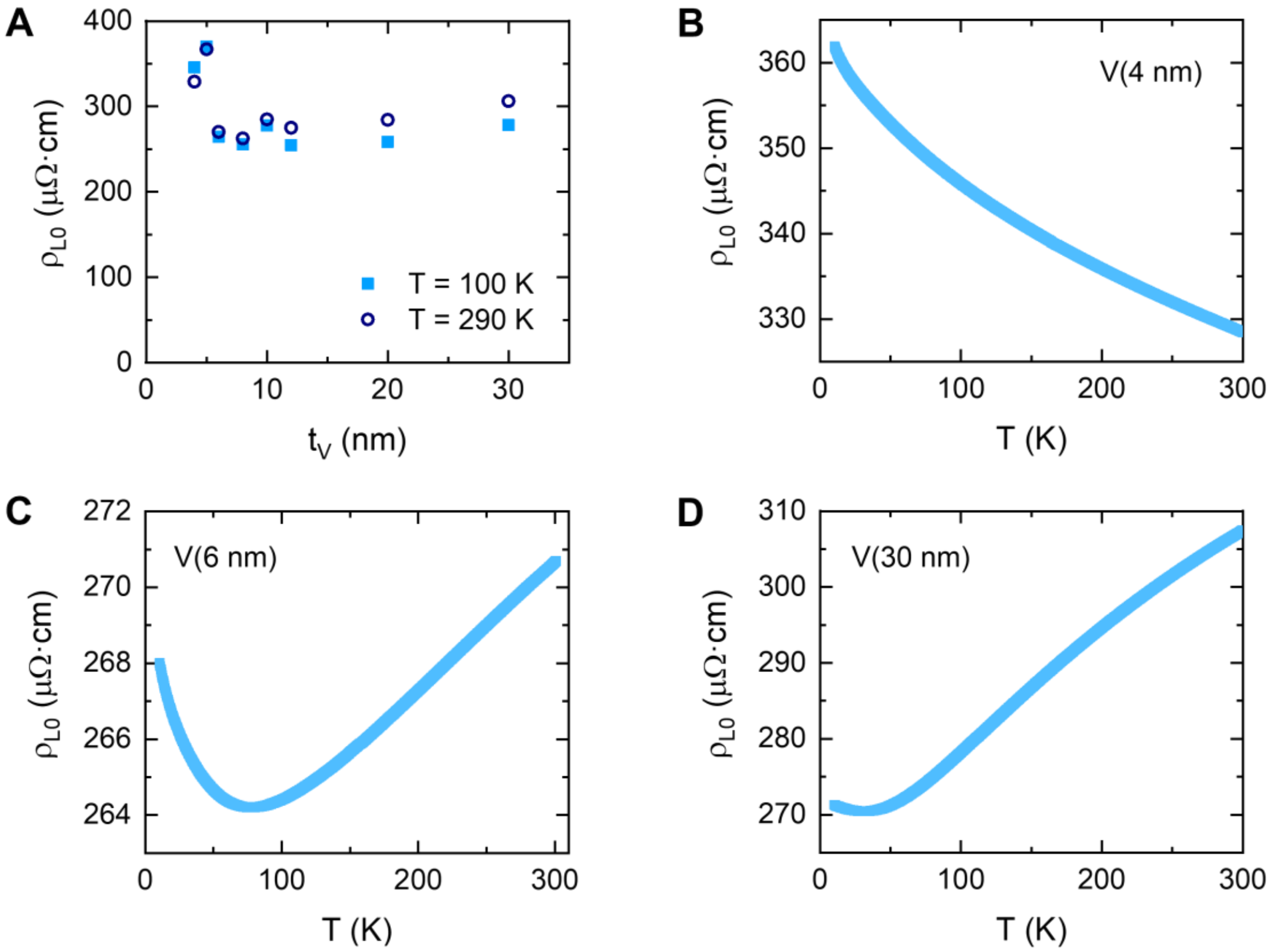

**Figure S1. Resistivity of V thin films.** (A) Resistivity of the set of V thin films with varying thickness, at 100 K and 290 K. (B-D) Resistivity of the 4-, 6-, and 30-nm-thick V films as a function of temperature.

**Note S2. Longitudinal magnetoresistance measurements**

Figure S2A shows characteristic longitudinal angle-dependent magnetoresistance (ADMR) measurements, performed in the sample Si/$SiO_2$/V(8 nm)/$SiO_2$(5 nm) along the three rotation planes $(\alpha, \beta, \gamma)$. The temperature dependence of each ADMR amplitude is plotted in Figure S2B.

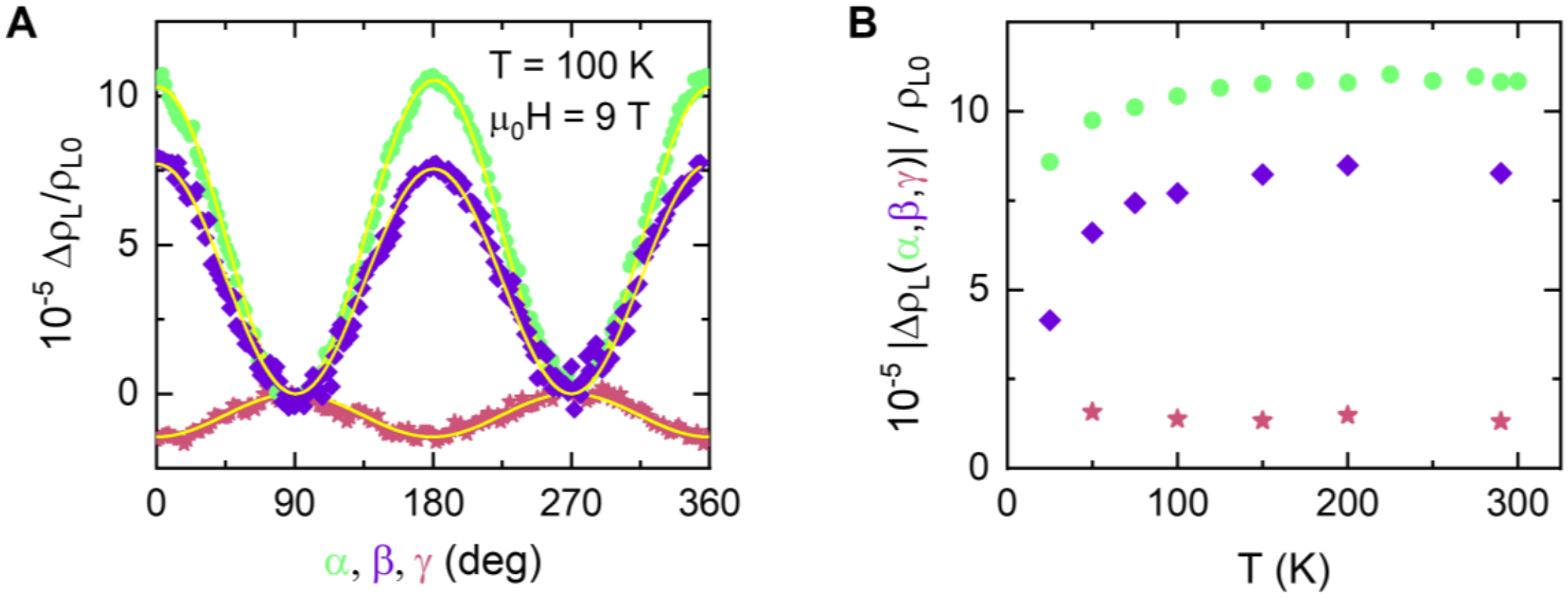

**Figure S2. Longitudinal angular-dependent magnetoresistance measurements, related to Figure 2.** (A) Longitudinal ADMR measurements at $\mu_0 H$ = 9 T and $T$ = 100 K along the three rotation planes ($\alpha$, $\beta$, $\gamma$). Yellow lines correspond to a $\cos^2(\alpha, \beta)$ and a $\sin^2(\gamma)$ fit. (B) Amplitude of the ADMR as a function of temperature. Data corresponds to the 8-nm-thick V film.

**Note S3. Transverse magnetoresistance measurements**

The transverse field-dependent magnetoresistance (FDMR) with the magnetic field $H$ applied along the three main axes ($x$, $y$ and $z$) shows a nonlinear relation with the magnetic field, with different slope at low and high magnetic fields. The linear trend at high fields comes from the ordinary Hall effect, expected for a nonmagnetic metal [1]. To extract the HMR contribution from the transverse FDMR, we perform a linear fit in the high field regime between -7 T and -9 T (see light blue data in Figure S3). Then, this linear fit is subtracted from the measured data (blue data in Figure S3) to eliminate the ordinary Hall contribution, and Figure 2D in the main text is obtained.

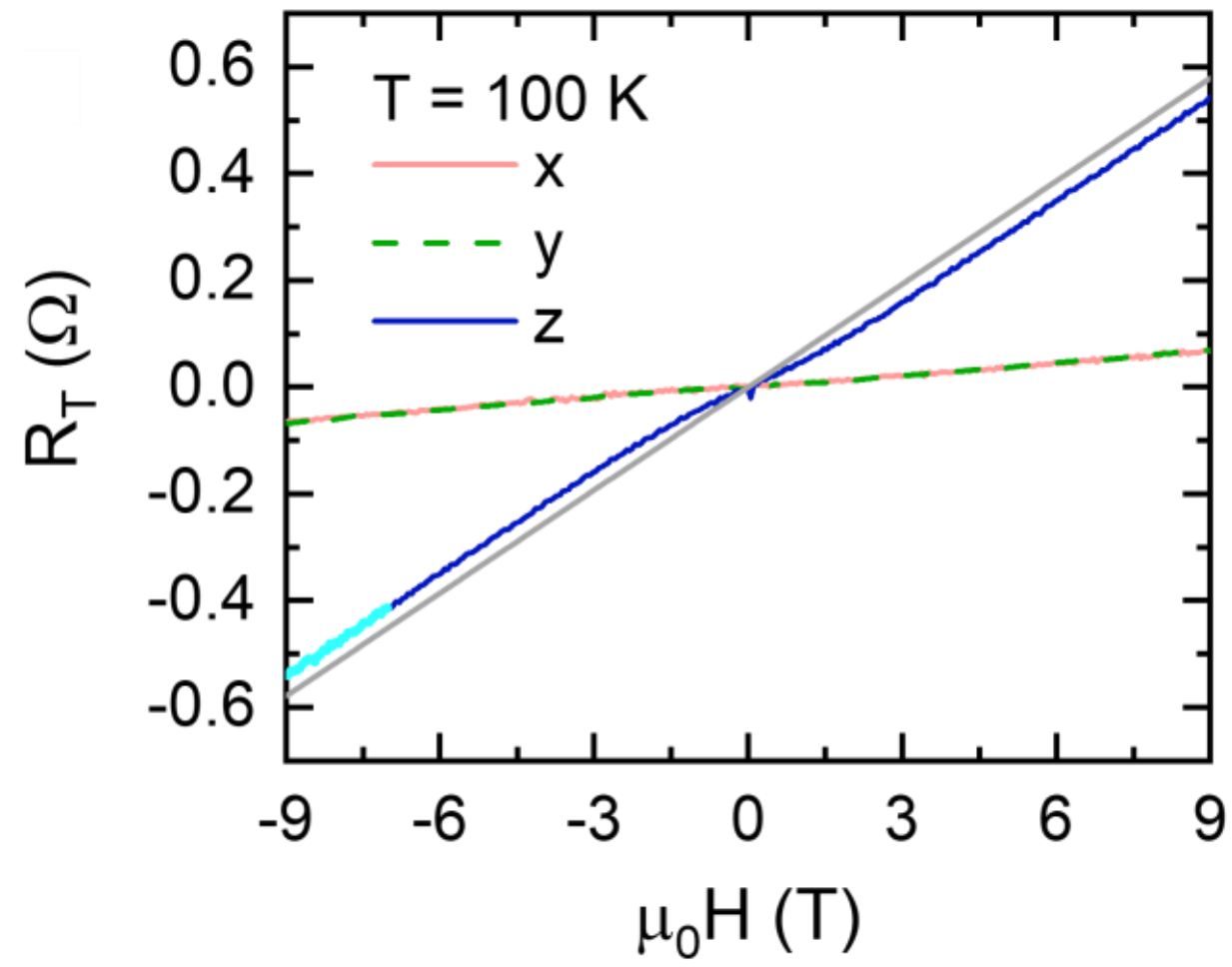

**Figure S3. Transverse field-dependent magnetoresistance measurements, related to Figure 2.** Measured transverse FDMR at $T$ = 100 K for the 6-nm-thick sample, with the magnetic field $H$ applied along the three main axes. The light blue data represents the region used to perform the linear fit when $H$ is along z. The grey line is the extrapolation of the linear fit that helps to visualize the change of trend between the low field and high field regime due to the ordinary Hall effect. A constant offset has been subtracted.

## Note S4. Hanle magnetoresistance of high-resistivity samples

As discussed in Note S1, V films with the same thickness but different resistivities (in the range $\rho_{L0}\sim$ 171–464 μΩ·cm) were obtained. We perform ADMR measurements in a set of samples (6-nm-thick and 10-nm-thick) with different resistivities at $\mu_0 H$ = 9 T and $T$=290 K (Figure S4A). A representative result of the ADMR, from a 6-nm-thick film with a resistivity of $\rho_{L0}$ = 378 μΩ·cm at 290 K, is plotted as $\Delta\rho_L/\rho_{L0} = \Delta R_L/R_L = [R_L(\alpha,\beta) - R_L(90°)]/R_L(90°)$ in Figure S4B. Moreover, the longitudinal and transverse FDMR show the characteristic HMR behavior described in the main text, as shown in Figures S4C and S4D.

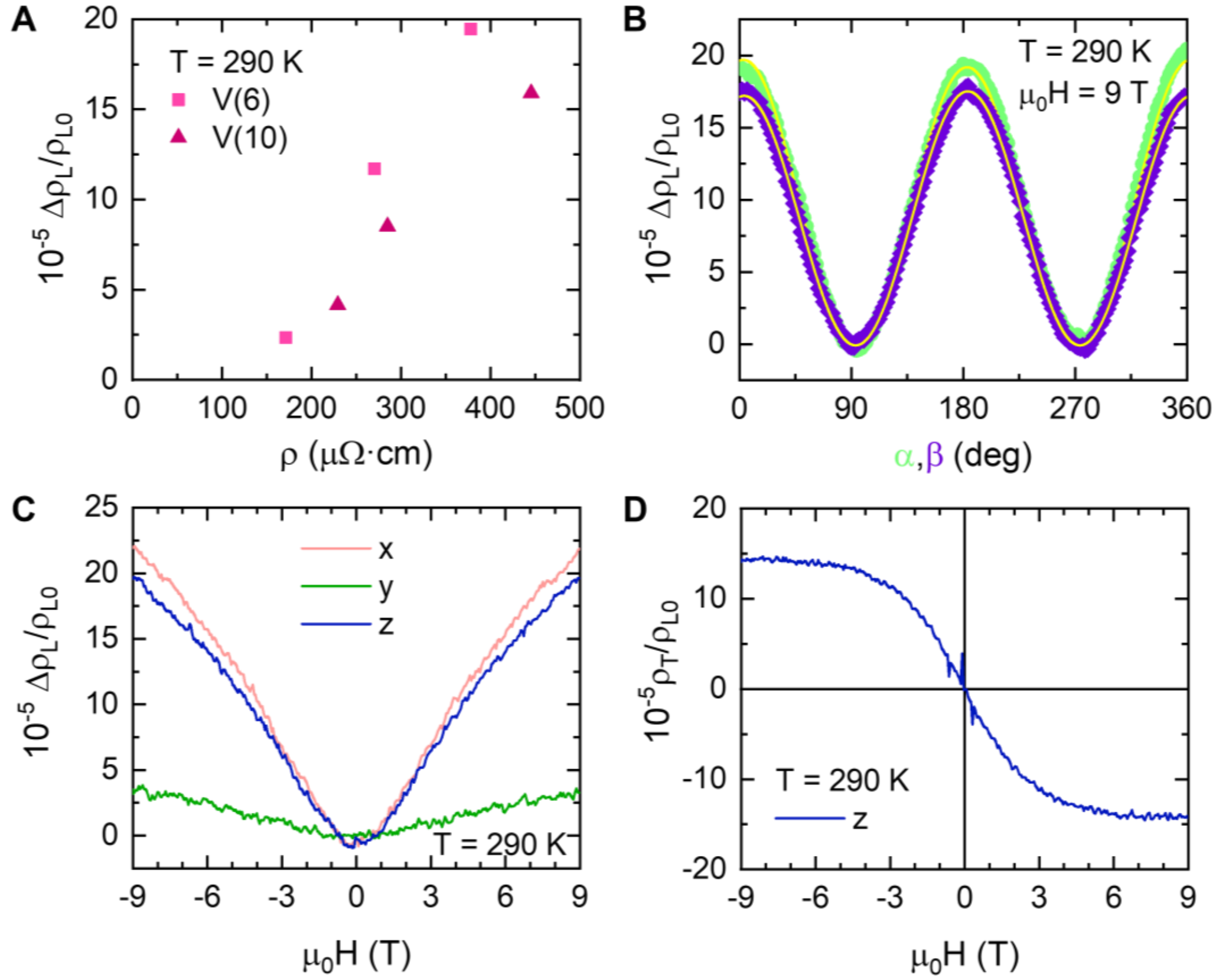

**Figure S4. Hanle magnetoresistance of samples with varying resistivity, related to Figure 2.** (A) Amplitude of the HMR taken from ADMR measurements, in $\alpha$ plane at $T$ = 290 K and $\mu_0 H$ = 9 T, as a function of the resistivity for different samples with V thicknesses of 6 and 10 nm. (B) Longitudinal ADMR measurements at $T$ = 290 K and $\mu_0 H$ = 9 T in $\alpha$ and $\beta$ planes. Yellow lines are a $cos^2(\alpha,\beta)$ fit. (C) Longitudinal and (D) transverse FDMR measurements performed at $T$ = 290 K with $H$ applied along the main axes. Data in (B), (C) and (D) correspond to sample Si/$SiO_2$/V(6 nm)/$SiO_2$(5 nm) with $\rho_{L0}$ = 378 μΩ·cm.

**Note S5. Orbital transport parameters of high-resistivity samples**

Taking into consideration the change of HMR amplitude in samples with high resistivity, we use Equations 3 and 4 from the main text to analyze the impact of increasing the resistivity in the orbital transport parameters. To do so, we perform a simultaneous fit with 3 free parameters ($\theta_{OH}$, $\lambda_{OD}$, and $D_O$) of the thickness dependence data set of high-resistivity samples ($\rho_{L0}$~378–464 μΩ·cm) and the longitudinal and transverse FDMRs of a 6-nm-thick sample with a resistivity of $\rho_{L0} = 378$ μΩ·cm, at 290 K (see Figure S5). This allows us to get the orbital transport parameters: $\theta_{OH}$, $\lambda_{OD}$, and $D_O$ shown in Table I of the main text.

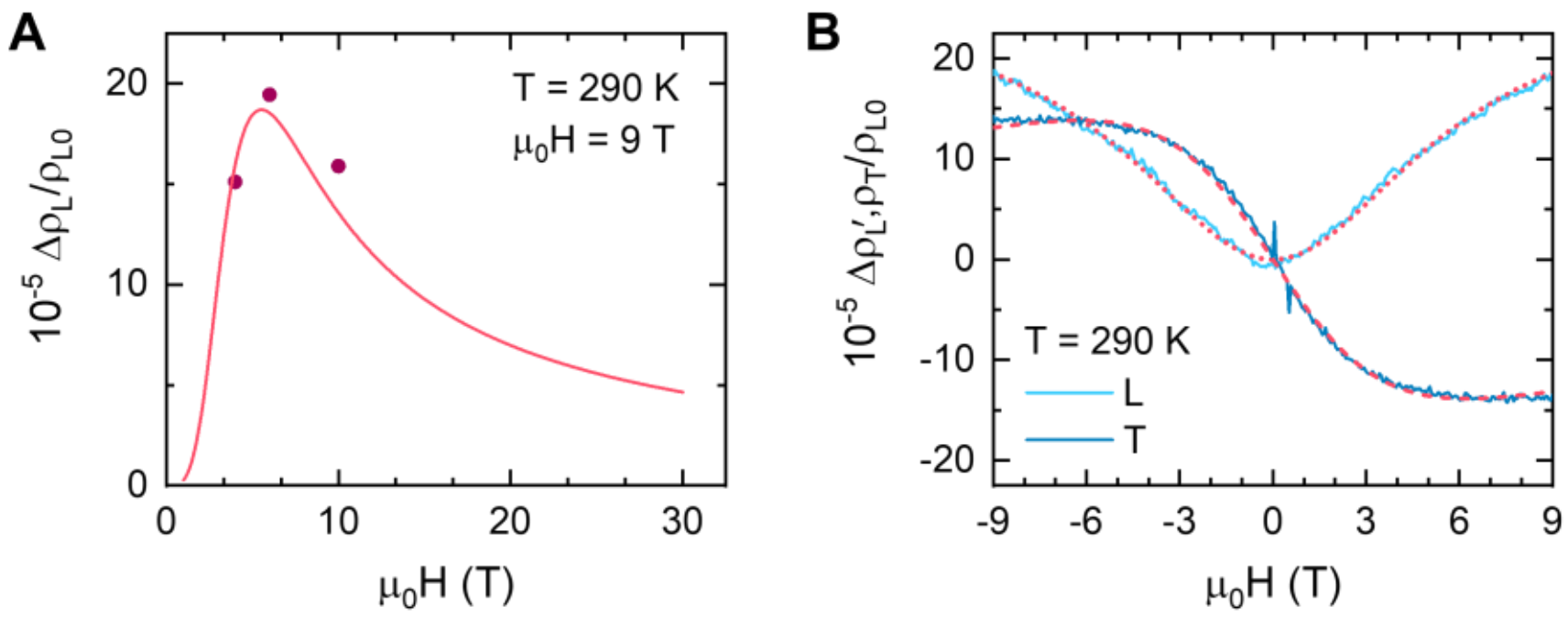

**Figure S5. HMR thickness and field dependence of high-resistivity V thin films with fittings to the HMR equations, related to Figure 3.** (A) Longitudinal HMR amplitude of high-resistivity samples at 9 T extracted from the ADMR measurements in $\alpha$ plane (pink solid circles) as a function of thickness at 290 K, with its corresponding fitting of Equation 3 (red solid line). (B) Longitudinal ($L$, light blue line) and transverse ($T$, dark blue line) FDMR at 290 K of the 6-nm-thick sample with high resistivity ($\rho_{L0} = 378$ μΩ·cm), with their respective fittings of Equation 3 (dotted lines) and Equation 4 (dashed lines) from the main text.

Furthermore, to assess the impact of the fitting method on the extracted orbital transport parameters ("Method 2"), we have also performed a triple-variable fit for the low-resistivity sample and the low-resistivity dataset (results are shown in Table S1). This change in fitting procedure does not significatively affect the extracted parameter values. The errors in Table S1 correspond to the statistical errors obtained from the fitting procedure, which, in the case of the fitting method described in the main text for the low-resistivity sample ("Method 1") cannot be quantified for certain parameters. The errors used in Table I of the main text correspond to an overall estimated error coming from several sources.

**Table S1. Orbital transport parameters of V extracted from the fittings at 290 K.** Method 1 refers to the fitting procedure where $\lambda_{OD}$ is first extracted using the thickness dependence dataset, followed by the fitting of the other two parameters ($\theta_{OH}$ and $D_O$) using the longitudinal and transverse FDMR datasets. Method 2 refers to the simultaneous triple-variable fitting approach using the three datasets.

| Method | $\rho_{L0}$ (μΩ·cm) | $\lambda_{OD}^{V}$ (nm) | $\theta_{OH}^{V}$ | $D_O^V$ (mm²/s) | $\tau_O^V$ (ps) | $\sigma_{OH}^{V}$ [(ħ/2e) Ω⁻¹cm⁻¹] |
|---|---|---|---|---|---|---|
| 1 | 270.3 | 2.1 | 0.0214 | 2.16 ± 0.02 | 2.04 ± 0.01 | 79 |
| 2 | 270.3 | 2.2 ± 0.1 | 0.021 ± 0.004 | 2.3 ± 0.1 | 2.16 ± 0.1 | 77 ± 1 |
| 2 | 378.1 | 2.09 ± 0.07 | 0.0240 ± 0.0003 | 2.00 ± 0.06 | 2.18 ± 0.07 | 63.5 ± 0.8 |

**Note S6. Magnetoresistance measurements on YIG**

As shown in the main text Si/$SiO_2$/V samples with thickness similar to YIG/V(7nm) shows a higher HMR amplitude (compare amplitudes in Figure 2C with Figure 4C of the main text), even though the samples have comparable resistivities ($\rho_{L0}^{YIG/V(7)} = 295$ μΩ·cm and $\rho_{L0}^{Si/SiO2/V(6)} = 264$ μΩ·cm at 100 K). This difference arises because of the fabrication process: while YIG/V(7nm) sample was fabricated using an etching process, Si/$SiO_2$/V samples were fabricated using a lift-off procedure. The change in fabrication method was chosen in order to have a clean interface in the YIG samples, which is crucially important in order to have a proper transmission of spins (i.e., large spin-mixing conductance) and, consequently, a reliable SMR signal. However, this change of fabrication method results in a reduced HMR signal for the YIG/V samples.

To clarify this point and demonstrate that the observed difference is due to the fabrication method and not to the resistivity or significant changes in the microstructure that could have affected the amplitude of the signal, we show in Figure S6 a sister sample deposited and fabricated at the same time, but on a Si/$SiO_2$ substrate ($\rho_{L0}^{Si/SiO_2/V(7)} = 302$ μΩ·cm). This means that both samples were sputtered and ion-beam etched simultaneously. As shown in Figure S6, both samples exhibit similar ADMR amplitudes, but lower than those obtained with the lift-off procedure. The different fabrication method may have introduced some resist residues or solvent degradation that affected the signal. In any case, it is important to note that there is still a difference of more than one order of magnitude between the SMR (Figure 4B) and HMR (Figure 4C and Figure S6) signals.

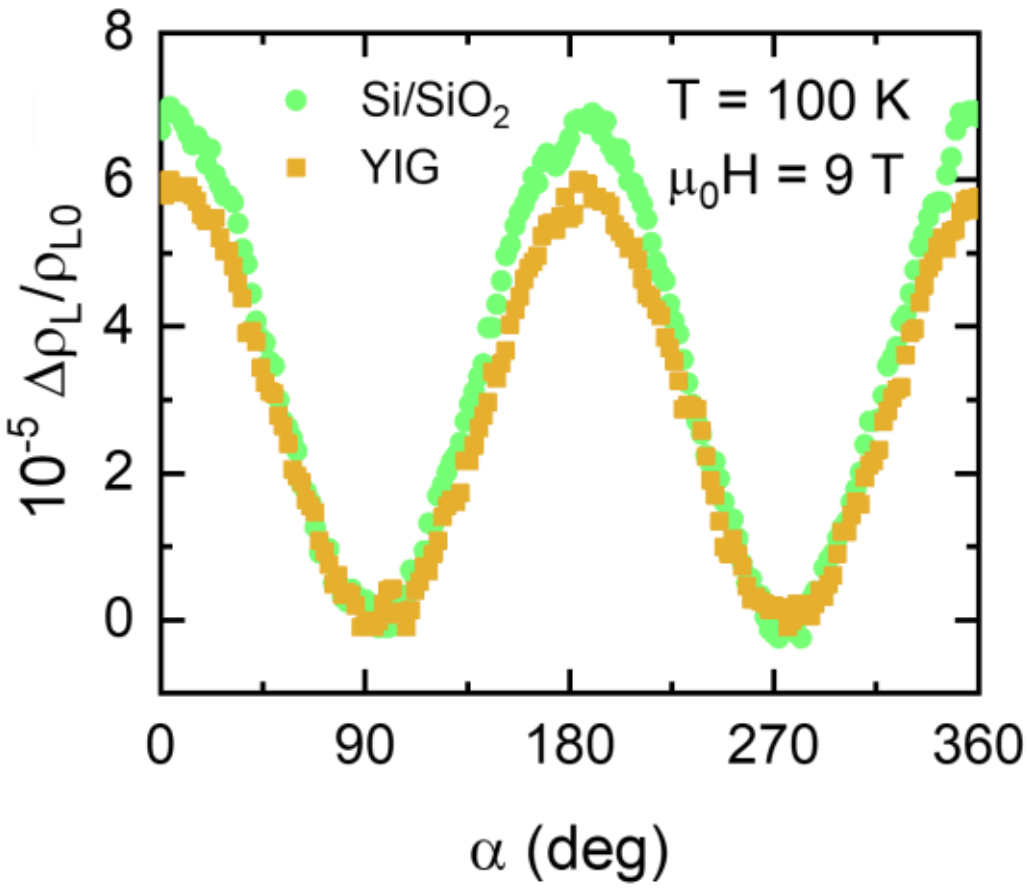

**Figure S6. ADMR measurements in V on different substrates, related to Figure 4.** Longitudinal ADMR measurements at $T$ =100 K and $\mu_0 H$ = 9 T in $\alpha$ plane for V(7nm) deposited on two different substrates: Si/$SiO_2$ and YIG. The two devices are processed with negative lithography and etching.

## Note S7. Structural characterization of V samples

We have performed X-ray diffraction (XRD) and reflectivity (XRR) to know the degree of crystallinity/disorder and thickness of the vanadium films, as well as atomic force microscopy (AFM) to quantify their roughness. Figure S7A shows the XRD patterns of Si/$SiO_2$/V($t_V$ nm)/$SiO_2$(5 nm) thin films, with $t_V$ = 5 and 30 nm, deposited together with the Hall bar samples used for magnetoresistance measurements. The absence of vanadium diffraction peaks indicates that the thin films have a very low crystallinity or are even amorphous, confirming their disordered nature. Figure S7B presents the corresponding XRR measurements, used to calibrate the thickness of the films. The similar decay in intensity observed for both thicknesses suggests that the films have comparable roughness. From the AFM scans, shown in Figures S7C and S7D, the roughness mean square (r.m.s.) of the thin films are estimated to be around 0.5 nm in both cases, corresponding to the two ends of the thickness range used in the main text.

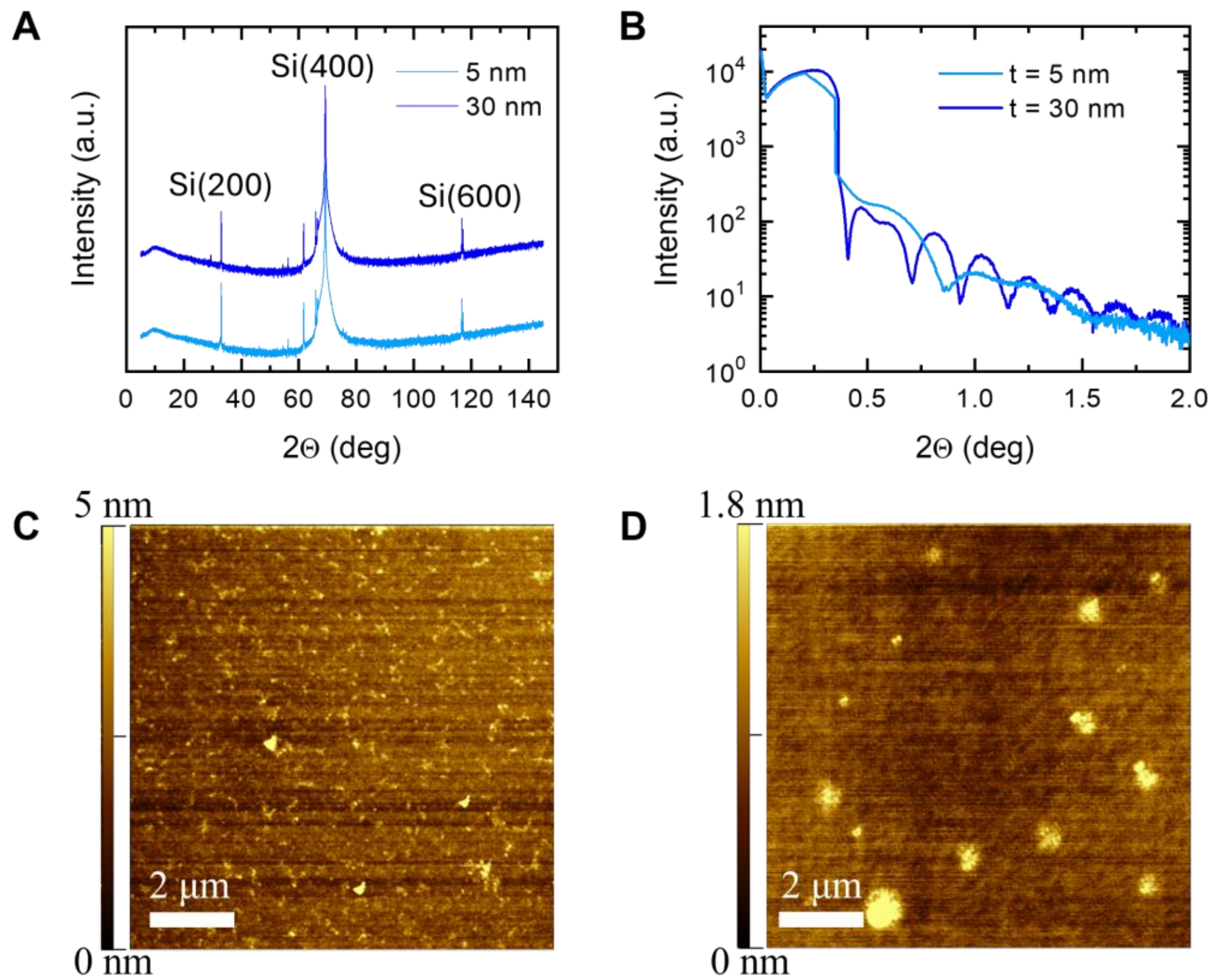

**Figure S7. Structural characterization of V samples**. (a) XRD spectra, (b) XRR, and (c,d) AFM from sputtered thin films of Si/$SiO_2$/V($t_V$ nm)/$SiO_2$ with $t_V$=5 and 30 nm, respectively.

### Note S8. Orbital transport parameters of low-resistivity 8-nm-thick sample

To further investigate the robustness of the orbital transport parameters, we performed an additional fitting for a low-resistivity 8-nm-thick sample using "Method 1" (see Figure S8). The extracted orbital transport parameters, shown in Table S2, are consistent with those obtained for the 6-nm-thick sample with similar resistivity analyzed in the main text. Furthermore, both samples follow the same trend with temperature, reinforcing the robustness of the observed trends.

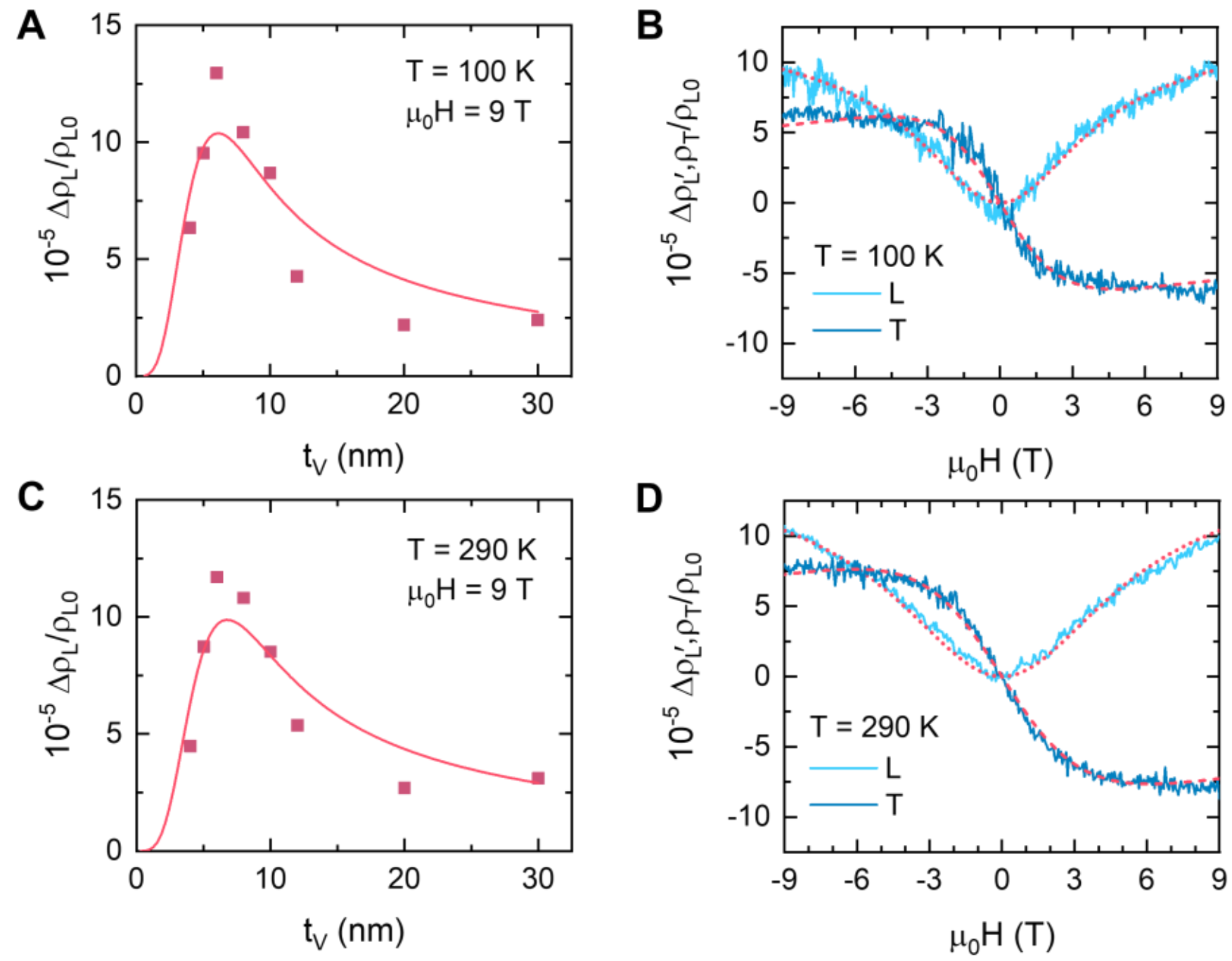

**Figure S8. HMR thickness and field dependence of low-resistivity V thin films with fittings to the HMR equations, related to Figure 3.** (A) and (C) HMR amplitude at 9 T extracted from the longitudinal ADMR measurements in $\alpha$ plane (pink solid squares) as a function of thickness at 100 K and 290 K, respectively, with their corresponding fittings of Equation 3 (red solid line). (B) and (D) Longitudinal ($L$) (light blue line) and transverse ($T$) (dark blue line) FDMR for sample Si/$SiO_2$/V(8 nm)/$SiO_2$(5 nm) at 100 K and 290 K, respectively, with their corresponding fittings of Equation 3 (dotted lines) and Equation 4 (dashed lines).

**Table S2. Orbital transport parameters of V extracted from the fittings at 100 K and 290 K for sample Si/$SiO_2$/V(8 nm)/$SiO_2$(5 nm).**

| T (K) | $\rho_{L0}$ (μΩ·cm) | $\lambda_{OD}^{V}$ (nm) | $\theta_{OH}^{V}$ | $D_{O}^{V}$ ($mm^2$/s) | $\tau_{O}^{V}$ (ps) | $\sigma_{OH}^{V}$ [($\hbar$/2e) $\Omega^{-1}cm^{-1}$] |
|---|---|---|---|---|---|---|
| 100 | 255.9 | 1.8 ± 0.3 | 0.019 ± 0.001 | 1.51 ± 0.5 | 2.14 ± 0.2 | 74 ± 4 |
| 290 | 262.7 | 2.1 ± 0.3 | 0.020 ± 0.001 | 2.46 ± 0.8 | 1.79 ± 0.2 | 76 ± 4 |

### Supplemental references

[1] J. Li, A. H. Comstock, D. Sun, and X. Xu, Comprehensive demonstration of spin Hall Hanle effects in epitaxial Pt thin films, Phys Rev B **106**, 184420 (2022).